\documentclass[fleqn,10pt]{wlscirep}

\usepackage[utf8]{inputenc}
\usepackage[T1]{fontenc}
\title{Noise Reduction in X-ray Photon Correlation Spectroscopy with Convolutional Neural Networks Encoder-Decoder Models }

\author[1]{Tatiana Konstantinova}
\author[1]{Lutz Wiegart}
\author[1]{Maksim Rakitin}
\author[1, +]{Anthony M. DeGennaro}
\author[1, *]{Andi M. Barbour}
\affil[1]{Brookhaven National Laboratory, NSLS-II, Upton, NY 11973, USA}

\affil[+]{adegennaro@bnl.gov}
\affil[*]{abarbour@bnl.gov}

\begin{abstract}
 Like other experimental techniques, X-ray Photon Correlation Spectroscopy is subject to various kinds of noise. Random and correlated fluctuations and heterogeneities can be present in a two-time correlation function and obscure the information about the intrinsic dynamics of a sample. Simultaneously addressing the disparate origins of noise in the experimental data is challenging. We propose  a computational approach for improving the signal-to-noise ratio in two-time correlation functions that is based on Convolutional Neural Network Encoder-Decoder (CNN-ED) models. Such models extract features from an image via convolutional layers, project them to a low dimensional space and then reconstruct a clean image from this reduced representation via transposed convolutional layers. Not only are ED models a general tool for random noise removal, but their application to low signal-to-noise data can enhance the data’s quantitative usage since they are able to learn the functional form of the signal. We demonstrate that the CNN-ED models trained on real-world experimental data help to effectively extract equilibrium dynamics’ parameters from two-time correlation functions, containing statistical noise and dynamic heterogeneities. Strategies for optimizing the models’ performance and their applicability limits are discussed.
\end{abstract}

\maketitle

\begin{document}

\section*{Introduction}
Noise reduction in experiments facilitates reliable extraction of useful information from a smaller amount of data. This allows for more efficient use of experimental and analytical resources as well as enables the study of systems with intrinsically limited measurement time, e.g. cases with sample damage or out-of-equilibrium dynamics. While instrumentation development and optimization of experimental protocols are crucial in noise reduction, there are situations where computational methods can advance the improvements even further.

\begin{figure}[!htb]
\centering
\includegraphics[]{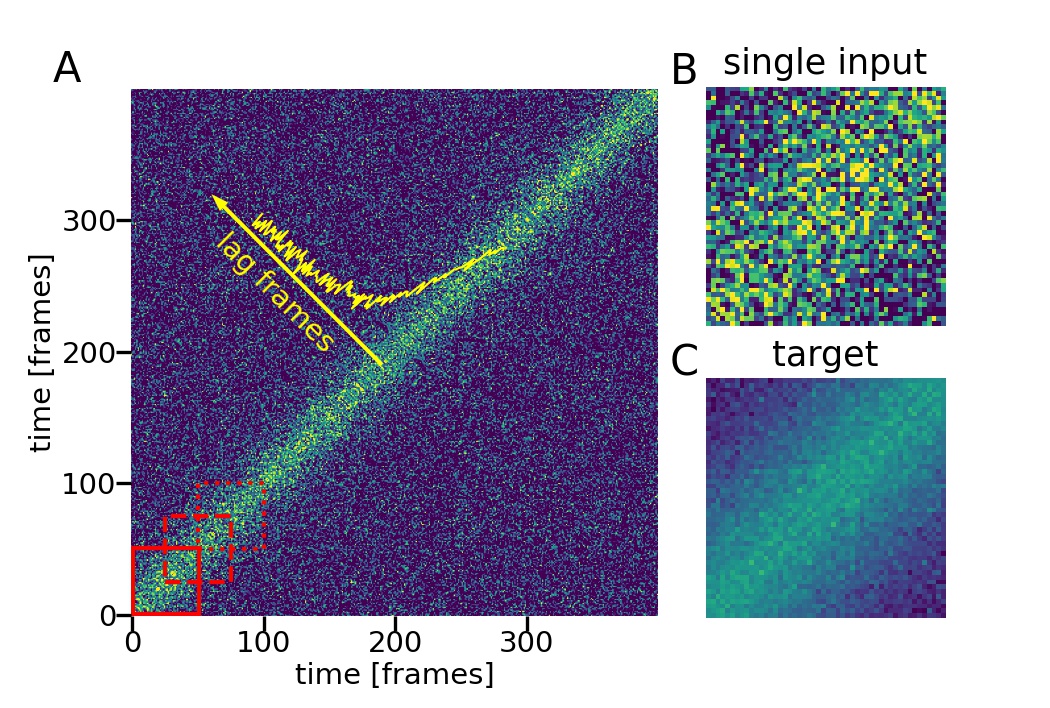}
\caption{Data for the model. (A) 2TCF for an experimental series consisting of 400 frames. Red squares show examples of regions selected for the model training. Yellow arrow shows the temporal direction \emph{t} of the system's dynamics. Yellow solid line shows the 1TCF along \emph{t}, calculated from the 2TCF. (B) Example of 50$\times$50 2TCF, passed as an input to the model. (C) Example of the target data for the model, obtained by averaging multiple 50$\times$50 diagonal sections of the 2TCF. All images have the same intensity scale.}
\label{fig:Figure1}
\end{figure}

\noindent X-ray Photon Correlation Spectroscopy (XPCS) \cite{Madsen_Fluerasu_Ruta, Shpyrko_2014, Sinha_2014} is a statistics-based technique that extracts information about a sample’s dynamics through spatial and temporal analysis of intensity correlations between sequential images (frames) of a speckled pattern collected from coherent X-ray beam scattered from the sample. The two-time intensity-intensity correlation function \cite{Brown_1997, Madsen_2010} (2TCF) is a matrix calculated as: 
\begin{equation}\label{eq:(1)}
C2(\pmb{q},t_{1}, t_{2}) = \frac{\langle I(\pmb{q},t_{1})I(\pmb{q},t_{2})\rangle}{\langle I(\pmb{q},t_{1})\rangle \langle I(\pmb{q},t_{2})\rangle}
\end{equation} 
where \(I(\pmb{q},t)\) is the intensity of a detector pixel corresponding to the wave vector \(\pmb{q}\) at time \(t\). The average is taken over pixels with equivalent \(\pmb{q}\) values.
An example of a 2TCF is shown in Fig.~\ref{fig:Figure1}. The dimensions of the matrix are \emph{N}x\emph{N}, where \emph{N} is a number of frames in the experimental series. The dynamics can be traced along the lag times \(\delta t=|t_{1}-t_{2}|\). In the case of equilibrium dynamics, information from a 2TCF can be ‘condensed’ to a single dimension by integrating along the \emph{(1,1)} diagonal producing a time-averaged one-time photon correlation function (1TCF) \cite{Luxi_Li_2014}: 
\begin{equation}\label{eq:(2)}
C1(\pmb{q},\delta t) = C_{\infty} + \beta|f(\pmb{q},\delta t)|^2
\end{equation}
where \(f(\pmb{q},\delta t)\) is the intermediate scattering function at lag time \(\delta t\), \(\beta\) is the optical contrast and \(C_{\infty}\) is the baseline that equals to 1 for ergodic samples.  While 1TCF can be directly obtained from raw data \cite{Lumma_2000}, calculating 2TCF as an intermediate step is beneficial even for  presumably equilibrium cases. 2TCF contains time-resolved information about both samples' intrinsic dynamics and fluctuations of the experimental conditions, which enables one to determine between stationary and non-stationary dynamics and whether or not the time-averaged 1TCF is a valid representation of the scattering series. Investigation of 2TCF helps to identify single-frame events, such as cosmic rays detection, and beam-induced dynamics, where timescales might vary with the accumulation of X-ray dose absorbed by the sample during the acquisition of the dataset. 

\noindent XPCS experiments can suffer from various sources of noise and artifacts: probabilistic nature of photon scattering, detector shot noise, and instrumentational instabilities. Significant progress in reduction of the noise involved in photon detection and counting has been made by developing single-photon counting devices \cite{Grybos_2016, Llopart_2002} and employment of the ‘droplet’ algorithm \cite{Livet_2000} or pixel binning \cite{Falus_2006}. Efforts have been dedicated to integrating feedback loops \cite{Kongtawong_2020, Strocov_2010} into instrumentational controls to reduce the impact of instabilities. Despite the current advances of experimental setup and methods for data analysis in reduction of noise and instability effects, achieving high signal-to-noise ratio is still a practical challenge in many XPCS experiments. The need to suppress the high-frequency fluctuations leads to extended data collection times – an approach that itself can introduce additional errors, for instance due to slow changes in experimental conditions. Limited experimental resources may not allow for multiple repeated measurements for systems with very slow dynamics. Besides, a sample’s intrinsic properties can limit the time-range, within which the dynamics can be considered \cite{Madsen_2010} as equilibrium and thus quantitatively evaluated with Eq.~\ref{eq:(2)}. A tool that helps to accurately extract parameters of the system’s equilibrium dynamics from a limited amount of noisy data would be useful, but no generally applicable, out-of-the-box tool exists for XPCS results.

\noindent Solutions based on artificial neural networks are attractive candidates as they are broadly used for application-specific noise removal. Among such solutions are extensions of autoencoder models \cite{ Kramer_1991}, which are unsupervised algorithms for learning a condensed representation of an input signal.  The principle behind an autoencoder is based on a common fact that the information about significant non-random variations in data is contained in a much smaller number of variables than the dimensionality of the data. An autoencoder model consists of two modules: encoder and decoder. The encoder transforms the input signal to a set of unique variables called latent space. The decoder part then attempts to transform the encoded variables back to the original input.  As the number of components in the latent space is generally much smaller than the number of components in the original input, the nonessential information, i.e. random noise, is lost during such transformations. Thus, an autoencoder model on its own can be used as an effective noise reduction tool. However, in the scope of this work we employ a broader idea of noise. We treat all dynamic heterogeneities due to changes in a sample configuration caused by stress or diffusive processes, as well as correlated noise in 2TCF, as an unwanted signal. Such point of view can be preferred when one wants to quantify the average dynamics parameters with Eq.~\ref{eq:(3)} or to separate the underlying (envelope) dynamics from stochastic heterogeneities. An autoencoder model can be modified to address the removal of a deterministic, application-specific noise by replacing its targets with  ‘noise-free’ versions of the input signals. In the case of an image-like input, such as an XPCS 2TCF, convolutional neural networks (CNN)  are the obvious choice for the encoder and decoder modules. CNN-based encoder-decoder (CNN-ED) models have been successfully implemented for noise removal and restoration of impaired signals in audio applications\cite{Grais_2017, Se_Rim_Park_2017} and images \cite{Pathak_2016, Xioa-Jiao_Mao_2016}.

\begin{figure}[!htb]
\centering
\includegraphics[]{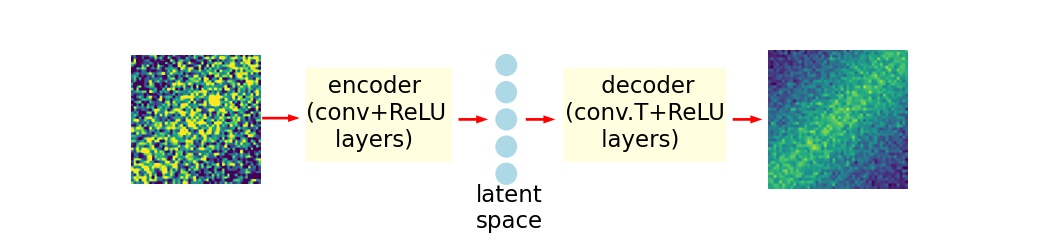}
\caption{ Architecture of the CNN-ED model. The input and the output images have the same intensity scale.}
\label{fig:Figure2}
\end{figure}

\noindent Here, we demonstrate an approach for noise reduction in 2TCFs by means of CNN-ED models. An ensemble of such models, trained on real experimental data, shows noticeable suppression of noise while preserving the functional form of system's equilibrium dynamics and the temporal resolution of the signal. Addressing noise removal from 2TCF instead of the scattering signal at the detector makes the approach agnostic to the type of the registering device, the size of the selected area, the shape of the speckles, the intensity of the scattering signal and the exposure time, enabling the models' application to a wide range of XPCS experiments.

\section*{Results}

\textbf{Data Processing.}
The models are trained using data from the measurements of equilibrium dynamics of  nanoparticle filled polymer systems conducted at the Coherent Hard X-ray Scattering (CHX) beamline at NSLS-II. For the nanoparticles’ dynamics Eq.~\ref{eq:(2)} can be approximated by the form \cite{Madsen_2010}:
\begin{equation}\label{eq:(3)}
C1(\pmb{q},t) = C_{\infty} + \beta e^{-2(\Gamma t)^\alpha}
\end{equation}
where \(\Gamma\) is the rate of the dynamics and \(\alpha\) is the compression constant. The baseline \(C_{\infty}\) is nearly 1 in the considered cases. Each experiment contains a series of 200-1000 frames. To augment the training data, additional 2TCFs are constructed using every second frame of the original series, which would be an equivalent to data collection with a twice longer lag period. Multiple regions of interest (ROI) - groups of pixels on the detector with equivalent wavevectors - are analyzed for each series and the 2TCFs are calculated for each ROI.
For each model datum, or an "example", the input image is obtained by cropping a 50x50 pixels part from a 2TCF with the center on the \emph{(1,1)} diagonal, starting at the lower left corner, as shown in Fig.~\ref{fig:Figure1}(A). Each next datum is obtained by shifting the center of the cropped image along the diagonal by 25 frames.
The target image for each example is the average of all the cropped inputs extracted from the same 2TCF. Thus, groups of 3 to 39 inputs have the same target. While the target images still contain noise, its level is significantly reduced with respect to the noise of the input images. Here, the size of 50x50 pixels is chosen as for the majority of the examples in the considered dataset the dynamics' parameters can be inferred from the first 50 frames. However, any size can be selected to train a model with little to no modification to its architecture if enough data are available. 

\begin{table}[ht]
\centering
\begin{tabular}{|l|l|l|l|}
\hline
 & Training & Validation \\
\hline

Unique Inputs & 12236 & 5449  \\
\hline

Unique Targets & 722 & 401 \\
\hline
\end{tabular}
\caption{\label{tab:Table1}Distribution of examples between the training and validation sets.}
\end{table}

\noindent The diagonal (lag=0) 2TCF values of the raw data reflect the normalized variance of the photon count. Such values are vastly different between experiments and detector ROIs. They can by far exceed the values of photon correlation between frames (typically on a scale between 1 and 1.5) and are usually excluded from the traditional XPCS analysis. To prevent the influence of the high diagonal 2TCF values on the model cost function, the pixels along the diagonal are replaced with the values randomly drawn from the distribution of 2TCF values at lag=1. In doing so, we avoid artificial discontinuities in the images.

\noindent For a proper model training process and to ensure its generalizability, we find that all the input data should be introduced to the model on the same scale. However, a commonly applied standard scaling is not suitable for the present case as the level of noise may affect the values of the essential parameters such as the baseline and the contrast. 
To bring all examples to a similar range, the estimated contrast for each series and each ROI is scaled to unity (see Methods). After processing, the data are split into the training and validation sets as shown in Table~\ref{tab:Table1}. The splitting is done in a way that no two inputs from different sets have the same target.

\noindent\textbf{Model Training.} The ED model architecture used in this work is shown in Fig.~\ref{fig:Figure2}. The encoder part consists of two convolutional layers with the kernel size 1$\times$1. Training the model with larger kernel sizes did not improve the performance of the model. While kernels of size 1$\times$1 are used sometimes in CNN image applications \cite{Simonyan_2014} for creating non-linear activations, generally, they are not exclusively incorporated across the entire network. The reason for this is that the convolutional kernels are intended to catch distinctive edges, which form characteristic features of an image. To identify an edge, the distribution of intensities among the neighboring pixels is needed.  However, the 2TCFs used in this work do not have sharp edges, which can partially explain the lack of improved learning with larger kernels. Besides, an equilibrium 2TCF has a unique  structure, with symmetry along the diagonals. An equilibrium 2TCF and its modified copy with pixel values randomly shuffled along the diagonals would produce exactly the same 1TCF. This property is picked up by the model during compression of convolutional outputs to the latent space.

\noindent Both convolutional layers consist of 10 channels with rectified linear unit (\emph{ReLU}) activation function applied to the output of each channel. 
We find that increasing the number of channels does not significantly change the performance of the model and the smaller number of channels gives poorer performance.
No pooling layers\cite{Nagi_2011} were introduced to prevent information loss\cite{Ronneberger_2015} at the encoding stage. The output of the convolutional layers contains 25,000 features. A linear transformation is performed to convert them to the latent space of a much smaller dimension. 
While some ED image applications implement fully convolutional architectures \cite{Xioa-Jiao_Mao_2016, Se_Rim_Park_2017}, we believe that the introduction of the linear layer for purpose of denoising equilibrium 2TCFs is beneficial. Not only does the bottleneck layer provide the regularization of the model, it also mixes the features derived by convolutional layers from different parts of the input image.
The decoder part consists of two transposed convolutional layers, symmetrical to the encoder part, that convert the latent space back to a 50$\times$50 image.
The \emph{ReLU} function is applied only to the output of the first decoder layer.

\noindent The mean squared error (MSE) between the denoised output and the target is a natural choice of cost function for many image denoising applications. The MSE is shown to be useful for image denoising even in cases of some noise being present in the target \cite{Lehtinen_2018}. Moreover,  presence of noise in the input data puts a regularization on model weights, enforcing contractive property \cite{Alain_2014} on the reconstruction function of denoising EDs. The goal of the model presented here is to reduce the noise in 2TCF in such a way that the 1TCF, calculated from the model output, is as close to the target 1TCF as possible. Thus, the model’s learning objective is modified by inclusion the MSE between respective 1TCFs into the cost function.

\noindent We find that the regularization, which is enforced by the noise in both inputs and targets, in conjunction with the early stopping based on the cost function for the validation set, is sufficient for the model to avoid over-fitting. Introducing additional weight regularization reduced the accuracy of the model, especially for the cases of fast dynamics.

\begin{figure}[!htb]
\centering
\includegraphics[]{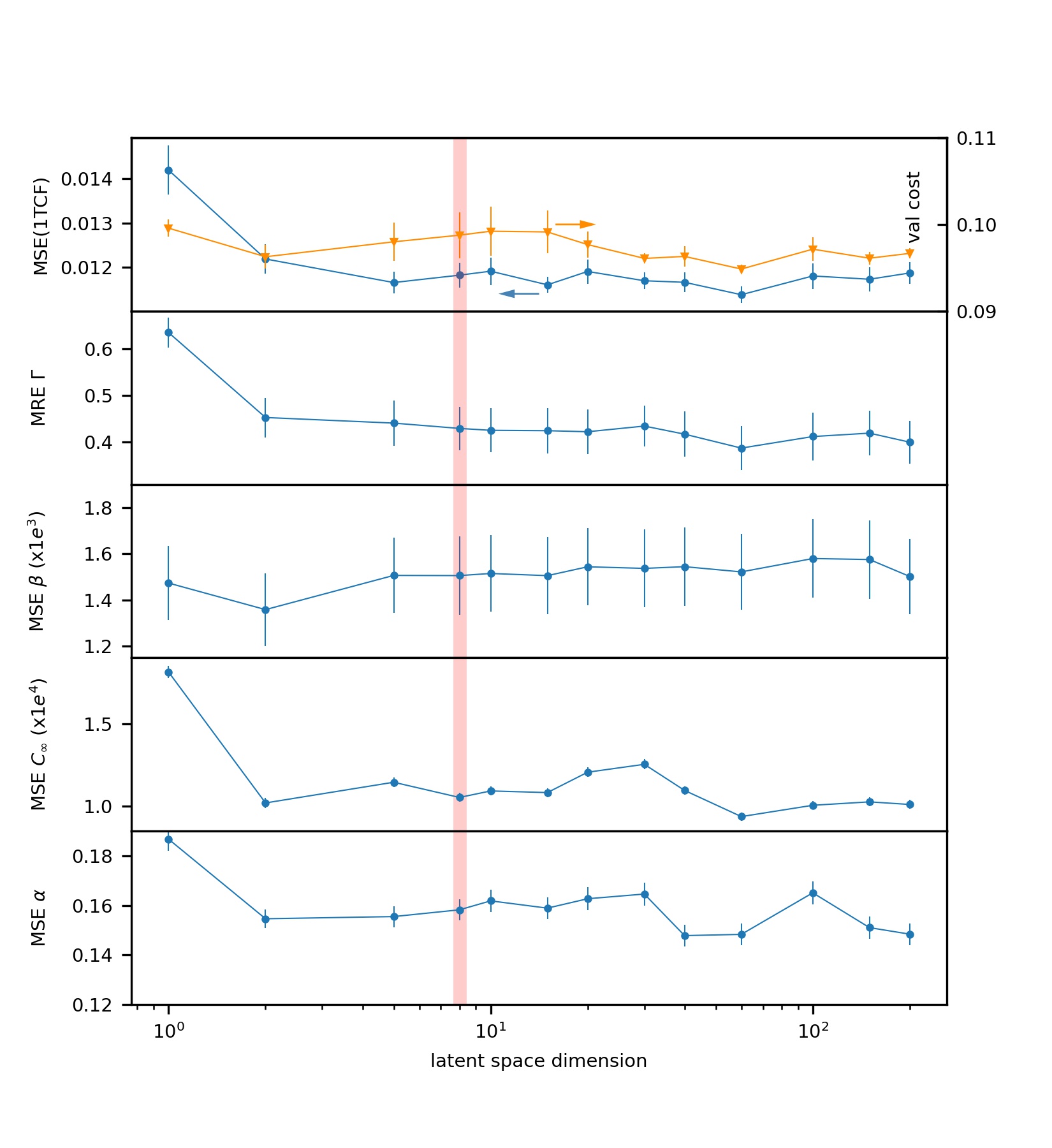}
\caption{Selection of the latent space dimension for the model. From top to bottom, as the function of the latent space dimension: the MSE(1TCF) (blue circles) and the cost function (orange triangles) for the validation set; the MRE($\Gamma$); the MSE of $\beta$, $C_{\infty}$ and $\alpha$, extracted from the fit of the corresponding CNN-ED ensemble's outputs for the examples in the validation set to Eq.~\ref{eq:(3)}. The vertical line marks the choice of the latent space dimension for the model.}
\label{fig:Figure3}
\end{figure}

\noindent However, the cost function calculated for the validation set is not the only parameter to consider when selecting the optimal parameters for the model.
When examining models trained for different latent space dimensions, the validation cost function (Fig.~\ref{fig:Figure3}) does not have a pronounced minimum in the range of dimensions between 1 and 200.

However,  this metric may not reflect well the systematic errors in reconstructing the
dynamics parameters, such as $\beta$, $\Gamma$, $\alpha$ and $C_{\infty}$, which are essential to drawing scientific conclusions. An efficient model would precisely recover these parameters for a broad set of observations. Thus, the optimal dimension is selected based on how  well the model output allows to recover those parameters for the validation data.
Here, the rate of the dynamics, $\Gamma$, is the most important parameter to consider since the variation of $\beta$ is taken care of by pre-processing normalization and the variations of $\alpha$ and $C_{\infty}$ are naturally very small in the considered examples.

\noindent To reduce the variance associated with the randomness of the initial weights initialization, ten models with different random initialization are trained for each latent space dimension. For each of the validation examples, the outputs of the ten models are converted to 1TCF, averaged and then fit to Eq.~\ref{eq:(3)}. The ground truth values, used for comparison, are obtained by fitting the 1TCF calculated from all (100-1000) frames in the same experiment and the same ROI as the input example is taken from.

\noindent Since values of dynamics rate can be very close to zero, the mean absolute relative error (MRE) is considered for $\Gamma$. The MSE is calculated for other parameters. The accuracy of $\Gamma$ keeps improving with increased number of hidden variables. But the rate of improvement slows down considerably above 5-8 variables. The same is observed for $\alpha$ and $C_{\infty}$. This is in agreement with the MSE(1TCF) between the model output and the target, as shown in Fig.~\ref{fig:Figure3}. The accuracy of $\beta$ is relatively uniform across all the models. Based on the above, we select the models with eight latent variables for further consideration.

\noindent To address the variance of the selected CNN-ED, we train 100 such models with different random initialization and select among them the 10 best performing models based on the MSE(1TCF) for the validation set. Selecting only a limited number of the best performing models instead of combining all trained models also optimizes the use of storage memory and computational resources.

\begin{figure}[!htb]
\centering
\includegraphics[]{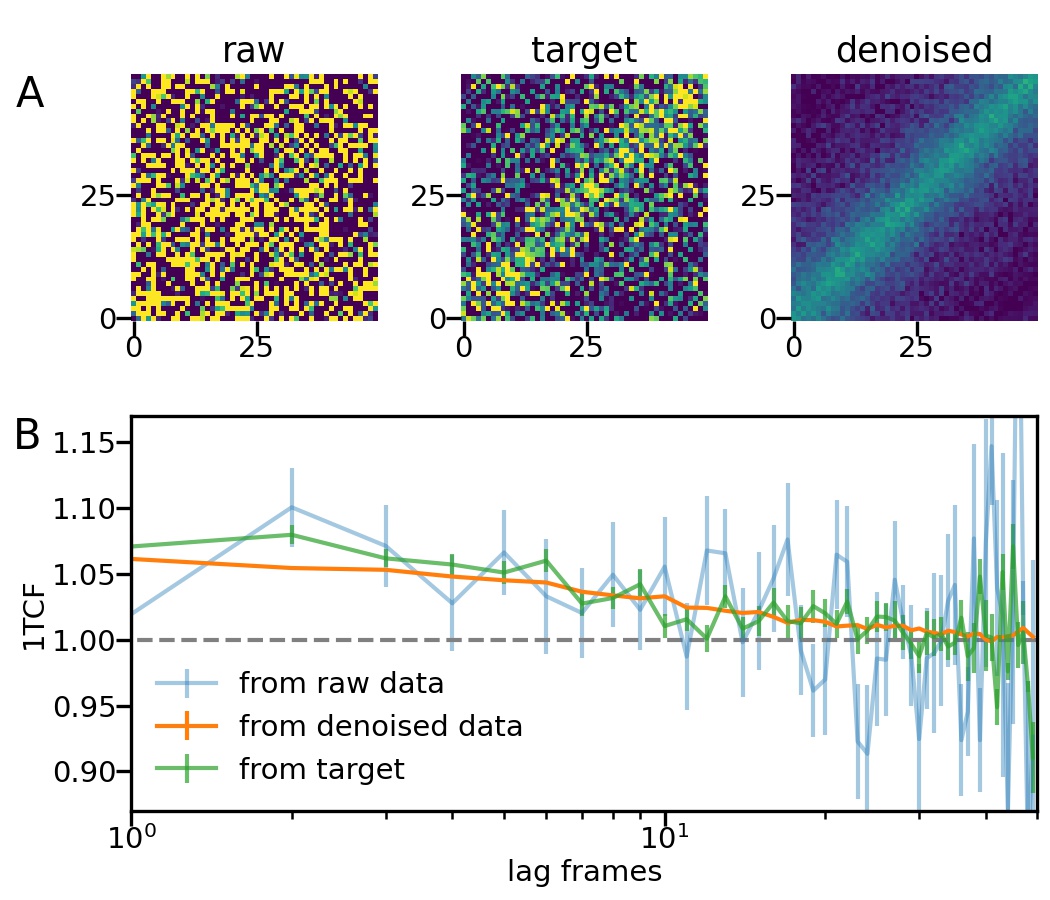}
\caption{Example of 2TCF denoising with the CNN-ED models. (A) From left to right: the raw input 2TCF; the averaged target; the output of the ensemble of CNN-ED models. (B) 1TCF calculated from each 2TCF in (A). The dashed line corresponding to a baseline \(C_{\infty} = 1\) is shown for convenience.}
\label{fig:Figure4}
\end{figure}

\noindent\textbf{Model Testing.} The performance of the ensemble of models is evaluated through several tests. Firstly, we estimate the model applicability range by applying it to experiments similar to the ones used for training.
An example of noise removal from a test datum is shown in Fig.~\ref{fig:Figure4}. Reduction of the noise is especially important for larger lag times, where fewer scattering frames are available for calculating the correlations. 

\noindent As mentioned above, despite the cost function working well for determining the optimal weights for a model, it is not sufficient to assess the reliability of the model output for quantitative analysis of the materials’ dynamics.  We assess the performance of the ensemble by comparing the fits with Eq.~\ref{eq:(3)} for the 1TCFs calculated from the cropped 50$\times$50 pixels raw data (inputs), the corresponding denoised model outputs and the full-range raw data (ground truth target) (see the Supplemental Materials). From the results of the test set, the noise removal from the raw cropped 2TCFs with the CNN-ED ensemble noticeably improves the precision for the dynamics parameters in a wide range of cases with $0.01 frames^{-1}<\Gamma<0.15 frames^{-1}$ (i.e. the contrast drops by half within approximately the first 3-35 frames) in comparison with fitting the raw cropped 2TCFs. The application of the model enables reasonable estimates even in cases when the low signal-to-noise ratio of the raw cropped data prevents a convergent fit within the parameters' boundaries. In the region $\Gamma >0.15 frames^{-1}$, the results of the model are no longer more accurate than the raw data in general. 
Note that the precision of the model depends on the accuracy of identifying the optical contrast. Accurate measurements of optical contrast in XPCS experiments with fast dynamics can be challenging as they can involve data collection with reduced exposure or relying on averaged speckle visibility.
Furthermore, a poor accuracy in identifying dynamics parameters is observed for inputs with very high noise levels (Fig.~S7) and/or the presence of well pronounced dynamical heterogeneities.

\noindent If 100 or more frames are available for analysis, 2TCFs with slow dynamics can be reduced by considering every $2^{nd}$, $3^{rd}$, etc. frame, as it is done for augmenting the training data. This to will effectively increase the exposure times and increase the $\Gamma$ measured in $frames^{-1}$, making the model output more accurate. Alternatively, a model with a larger size of input 2TCF can be trained to handle cases of slow dynamics.

\begin{figure}[!htb]
\centering
\includegraphics[]{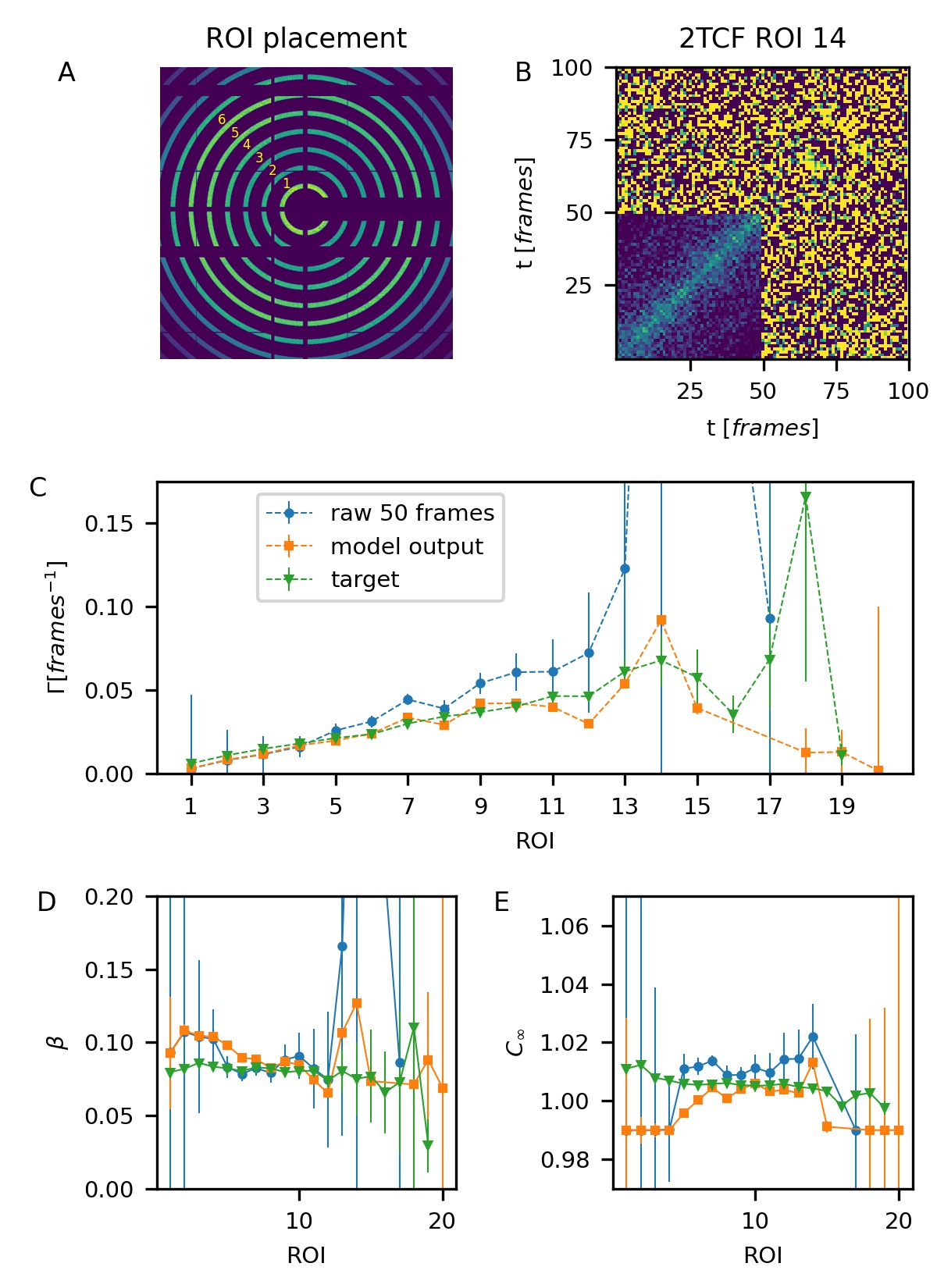}
\caption{Model application for recovering the $q$-dependence of the dynamics parameters. (A) Scattering image of the sample with the ROI map on top of it. Dark blue corresponds to pixels excluded from the analysis. (B) A 100-frame fragment of 2TCF from ROI \#14. The first 50-frame part is denoised with the model. Variation of $\Gamma$ (C), $\beta$ (D) and $C_{\infty}$ (E) parameters among ROIs with different $q$.}
\label{fig:Figure5}
\end{figure}

\noindent While it is clear from above how the model performs on average for individual independent 2TCFs, it is useful to see if application of the model can lead to reducing data collection in a typical XPCS experiment. We consider a single 700-frames series of scattering images among those used for creating the test set. The goal is to see if one can extract a sufficient information about the $q$-dependence of the dynamics rate $\Gamma$ using only the first 50 frames with and without the model application. The target 2TCF for each of the concentric ROIs (shown in Fig.~\ref{fig:Figure5}(A)) are calculated using all 700 frames. The first $50\times50$ frames regions of the 2TCFs are considered and the ensemble CNN-ED model is applied to them. The visual comparison between the level of noise in the raw data and in the model output for an ROI with large $q$ is shown in Fig.~\ref{fig:Figure5}(B). The 1TCFs, calculated from the raw cropped 2TCFs, from the model outputs and from the target 2TCFs for each ROI are fit to Eq.~\ref{eq:(3)} with $\alpha = 1$. The results are shown in Fig.~\ref{fig:Figure5}(C-E). For the parameter $\Gamma$, at small $q$, where the signal-to-noise ratio is high, all three fits are close. However, as $q$ grows and the noise level increases, the fit for the raw 50-frame 1TCFs starts to deviate from the target values more than the fit for the model outputs. In fact, the outcome of the model remains close to the actual values until the large $q$ values (ROI\# 16 and above). A similar tendency is observed for parameters $\beta$ and $C_{\infty}$. This example demonstrates that application of the model can help to obtain sufficient information about the equilibrium system's dynamics from a smaller amount of data.

\noindent The applicability of the model to non-equilibrium data is also tested. Although the model is trained with the equilibrium examples, it still can be applied to quasi-equilibrium regions of a 2TCF with gradually varying dynamics parameters. Here, the model performance is demonstrated for a sample with ageing dynamics that become slower with time. Since the target values cannot be obtained by averaging many frames for a such case, we calculate two 2TCFs with different noise levels, but carrying the same information, through sub-sampling pixels from the same ROI. The original ROI is a circle of small width with its center at $q$=0. This ROI is used for calculating the target 2TCF. To calculate the test 2TCF with the reduced signal-to-noise ratio we randomly remove 74\% of pixels from the original ROI. The model is applied along the \emph{(1,1)} diagonal in a sliding window fashion (see Methods).

\begin{figure}[!htb]
\centering
\includegraphics[]{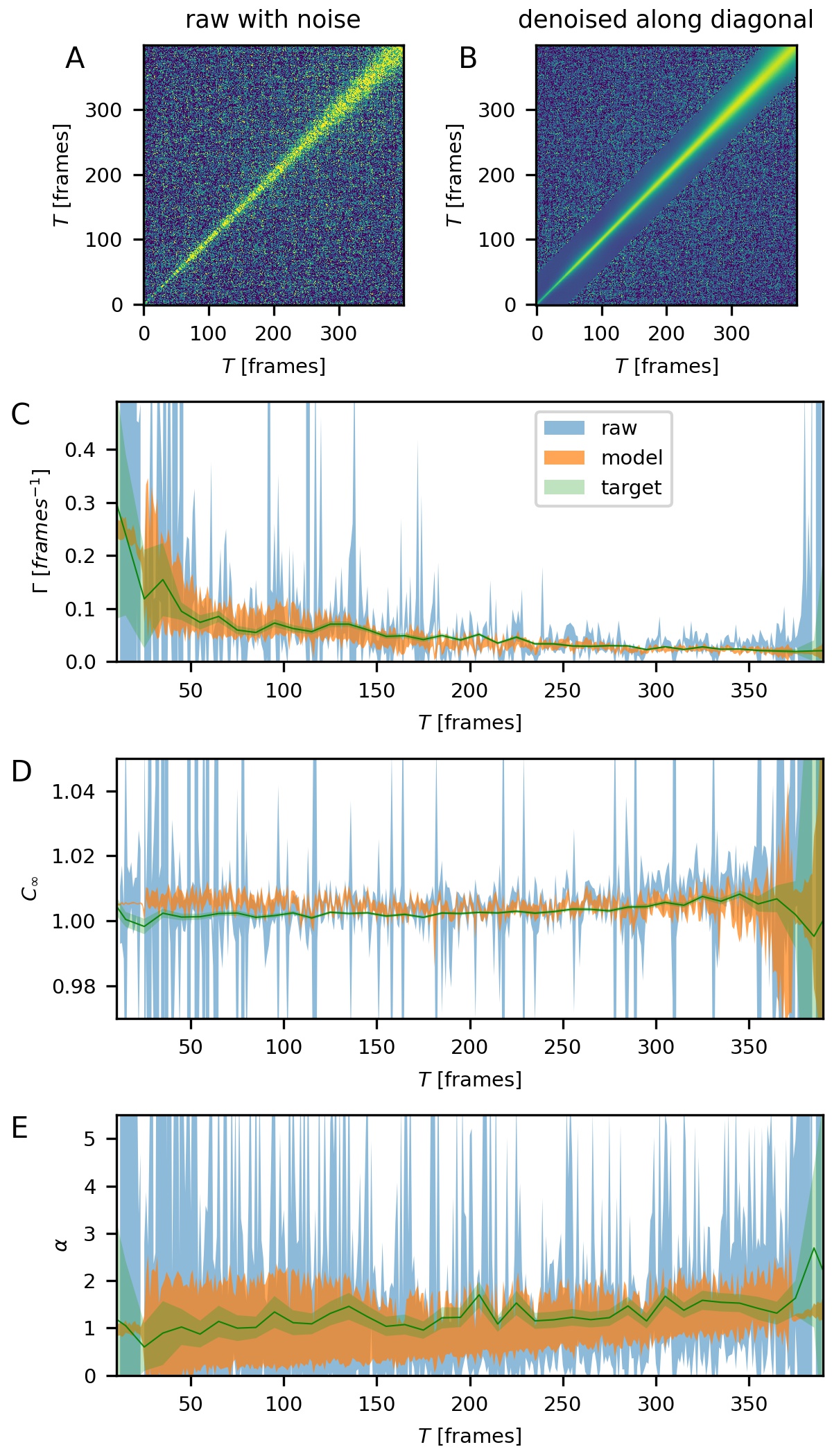}
\caption{Model application for the case of non-equilibrium dynamics. (A) 2TCF calculated from reduced ROI. (B) The same 2TCF after the model is applied along its diagonal. Temporal evolution of $\Gamma$ (C), $C_{\infty}$ (D) and $\alpha$ (E) parameters extracted from the noisy raw data, denoised data and the target 2TCF.}
\label{fig:Figure6}
\end{figure}

\noindent To compare the test 2TCF and the result of the model application, the cuts with width of 1 frame are taken perpendicular to the \emph{(1,1)} diagonal and the resulting 1TCFs are fit to Eq.~\ref{eq:(3)} in analogy to other XPCS analyses \cite{Madsen_2010, Malik_1998}. The target parameters are obtained by taking the cuts of 10 frames with the step of 10 frames from the target 2TCF and fitting the 1TCF, averaged over each cut, to the Eq.~\ref{eq:(3)}. Averaging is done for improving the accuracy of the target values. 
The contrast $\beta$ is estimated as the mean of the respective raw 2TCF(lag=1) at frames 250-300, where the dynamics are fairly slow, and is fixed during the fit.
The results for $\Gamma$, $C_{\infty}$ and $\alpha$ are shown in Fig.~\ref{fig:Figure6}(C-E). While the general trend of $\Gamma(t)$ could be visually estimated from the raw test data, the output of the model gives much fewer outliers. Moreover, the temporal region, where $\Gamma$ can be reasonably estimated is wider for the model output. The fit to the raw test data does not allow to estimate $\Gamma$ in the first 30-40 frames, while the fit to the denoised data is close to the target in that region. The fit of the denoised data only shows a high uncertainty at the corners of the 2TCF, where the corresponding 1TCFs consist of less than 20 points. The variance of parameter $C_{\infty}$ is also improved for the denoised data, but the most notable improvement in accuracy is observed for the parameter $\alpha$. The fits to the raw noisy data have high variance, which hides the upward trend of $\alpha$, in contrast to the fits to the denoised data. 

\noindent In a typical experiment, cuts with width of more than 1 frame are used for estimating the dynamics parameters achieving a better accuracy for the raw data than shown in Fig.~\ref{fig:Figure6}. However, the selection of regions with quasi-equilibrium dynamics is not trivial. Since the fits to 1-frame-wide cuts from the denoised data have a low variance almost across the entire experiment, application of CNN-EDs makes the data more suitable for automated analysis and for visual inspection of the data when selecting the quasi-equilibrium regions.

\begin{figure}[!htb]
\centering
\includegraphics[]{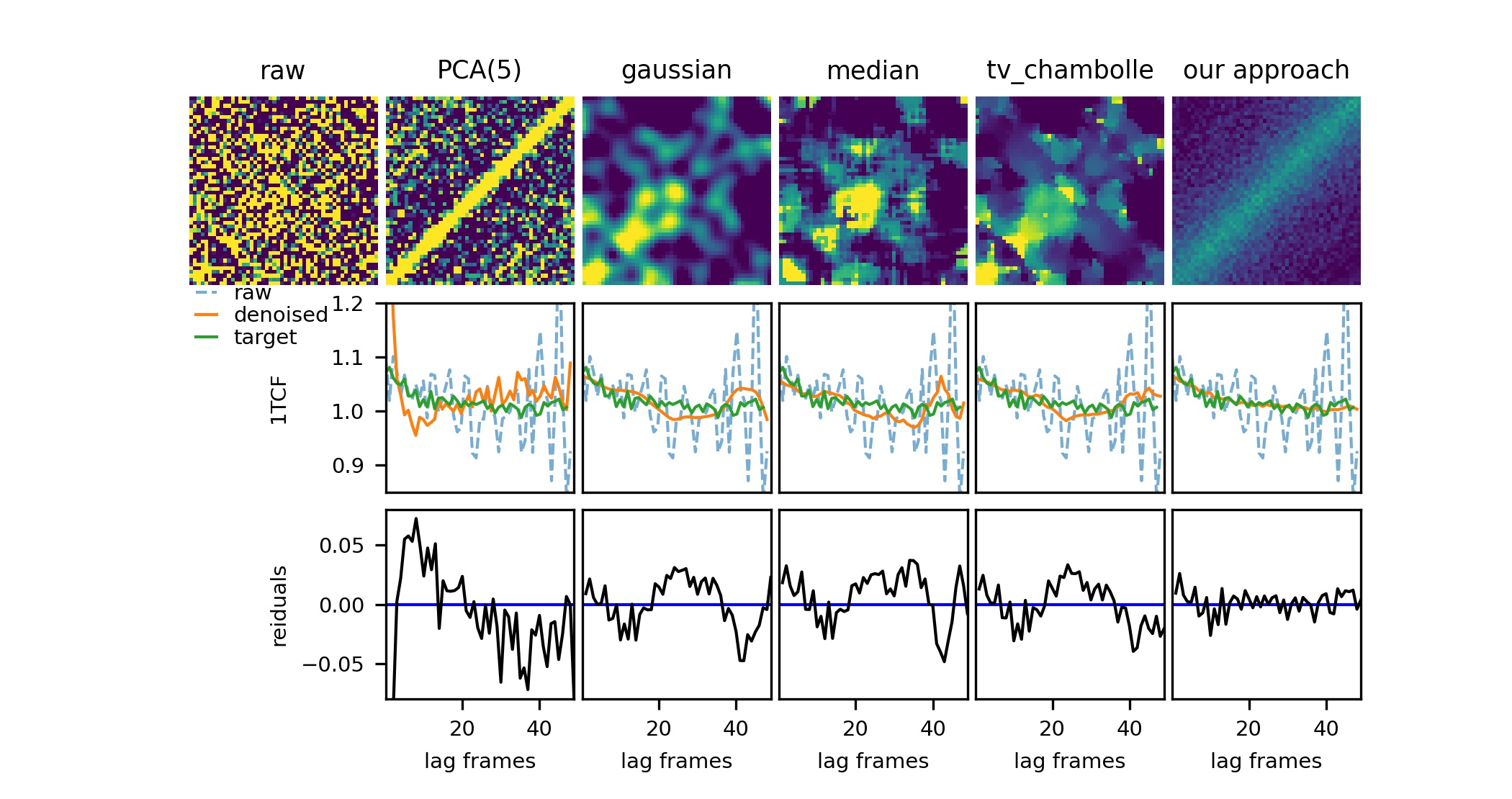}
\caption{Comparison of various noise removal techniques applied to an example from the test set. Top row: results of applying filters to the raw 2TCF, middle row: 1TCFs calculated from the 2TCFs for the raw input (blue dashed line), the results of the respective filters (orange solid line) and the target (green solid line), bottom row: residuals of the 1TFCs calculated from the example after denoising with the respective filters. }
\label{fig:Figure7}
\end{figure}

\noindent\textbf{Comparison to Other Techniques.} We compare the performance of our approach to several of-the-shelf solutions for noise reduction in images: linear principle components–based, Gaussian, median and total variation denoising (Chambolles’ projection) \cite{Duran_2013} filters. The comparison of the application of these techniques to the same test example as in Fig.~\ref{fig:Figure4} is shown in Fig.~\ref{fig:Figure7}. Principle components filters have the same idea as the ED model – preserving only the information from a few essential components of the original data. In fact, an autoencoder is a type of non-linear principle component generator. As one would expect, a filter based on linear principle components, trained on the same data as the CNN-EDs, under-performs comparing to the case of non-linear components due to a larger bias of the procedure for the components extraction. Gaussian and median filters are based on smoothing the intensity fluctuations between neighboring pixels and the total variation denoising is a regularized minimization of the additive normally distributed noise. While these approaches help to reduce pixel-to-pixel intensity variations, unlike the demonstrated here CNN-ED models, they do not learn the functional form of the equilibrium 2TCF images and cannot be improved by having a larger training set. By only considering local surrounding of individual pixels in a single image, such algorithms cannot recognize, for example, that the correlation function decays at larger lag times. Consequently, when an isolated high-intensity pixel (noise) is encountered in an image, an application of such filters leads to inflation of intensities in the surrounding pixels, highlighting the noise instead of correcting it. Thus, noise removal with the above filters can introduce false trends in 1TCF, which  makes them  unsuitable for quantitative XPCS data analysis. On the other hand, a CNN-ED, which is a regression model, learns from numerous examples the characteristic trends in the data and is less likely to introduce artifacts.

\section*{Discussion}
The CNN-ED approach to noise removal in XPCS shows a reasonable improvement in the quality of the signal, allowing for quantification of a sample’s dynamics from a limited amount of data, avoiding extensive data collection, accessing finite regions of reciprocal space and quasi-equilibrium intervals of non-equilibrium dynamics. The CNN-ED models go beyond and are superior to a simple smoothing of intensity between neighboring pixels since these models empirically learn the structural form of the 2TCF. The models are fast to train and do not require an extensive amount of training data. Their accuracy is pretty robust with respect to the choice of hyperparameters such as the number of channels in the hidden layers, the convolutional kernel size and the latent space dimension. The computational resources required for the application of the ensemble of 10 models are smaller than one needs to calculate 2TCFs for a typical number of frames required to achieve the same signal-to-noise ratio. 

\noindent However, there are several limitations to keep in mind when applying CNN-ED models to a 2TCF. The testing results show that the models may not reliably remove the noise for the cases of very fast and very slow dynamics as well as from very noisy data (see the illustration in Supplementary Fig.~\ref{fig:FigureS7}). Some inaccuracy for the cases of fast dynamics comes from uncertainties in identifying the normalization factors (contrasts) for pre-processing of the inputs, which is also a challenge for traditional analysis. When the speckle visibility drops significantly within a single frame acquisition period, its estimation from the input data can have a high error. As it is seen from the model performance for the validation set and for the non-equilibrium test case in comparison to its performance for the equilibrium test set, a more accurate scaling of the inputs can improve the precision of the model for experiments with faster dynamics.
Besides, one is advised to consider the context of extracted dynamics for a given material before relying solely on the information extracted from only a single 2TCF regardless of whether a CNN-ED model is applied.
One benefits from a series of experiments on a single system, such as a temperature dependence or the demonstrated here $q$-dependence, to lend credibility to extracted dynamics for one particular experiment.

\noindent In this work, only equilibrium dynamics described by stretched exponents with the baseline close to 1 are used for training. Thus, the model learns to approximate any input with this type of dynamics. This can result in a loss of fine details, such as heterogeneities, oscillations or fast divergence of dynamics parameters in non-equilibrium cases. However, the demonstrated approach to the noise removal can be expanded to other types of dynamics with sufficient amount of data for training. Even in the absence of proper denoised target data, the autoencoder version of the model can significantly reduce the random noise. Furthermore, a CNN-ED model can be trained to correct for specific types of artifacts, such as impact of instrumentational instabilities or outlier frames, leading to a more efficient use of experimental user facilities \cite{Campbell_2020}. Similarly to other fields\cite{Baur_2019, Chong_2017}, the autoencoder models can be used for identifying unusual observations in the stream of XPCS data. Additionally, the encoded low-dimensional representation of the 2TCF can be used for classification, regression and clustering tasks, related to samples’ dynamics. In the broader scope, the presented here CNN-ED models and their modifications have the potential for application in automated high-rates XPCS data collection\cite{Zhang_2021} and processing pipelines, reducing the reliance on the human-in-the-loop in decision making during experiments.  

\section*{Methods}

\noindent\textbf{Training data.} The data for training and validation set contain experiments for 7 samples from 3 different material groups A(1 sample), B(2 samples) and C(4 samples). The experiments are taken at various exposures, acquisition rates and temperatures. Concentric ROIs with increasing $q$ are used. Depending on the noise level and the dynamics duration, from 2 to 17 ROIs (median 10 ROIs) are considered for each experiment. The diversity of experimental conditions and regions in the reciprocal space allows one to obtain a realistic distribution of dynamics parameters and noise levels. We have not included the cases with very slow dynamics, for which only a small portion is complete within the 50 frames. To cut off the high noise data, we excluded the cases, where the fit to Eq.~\ref{eq:(3)} did not converge for the full-range 1TCF.  
The distributions of dynamics parameters for the training and the validation set are shown in Fig.~\ref{fig:FigureS1}.
For the model training purposes, all input data (2TCF$_{noisy}$) are scaled as:
\begin{equation}\label{eq:(4)}
input = (2TCF_{noisy} - 1)/\beta^{*} + 1
\end{equation}
where $\beta^{*}$ is the estimation of speckle visibility for the integration time of a single frame. It is obtained from fitting the equivalent pixels’ intensity fluctuations with a negative binomial distribution \cite{Luxi_Li_2014}. For this, the speckle visibility, is calculated for each frame and is averaged among all the frames in the series. The target data are reversely scaled accordingly.

\noindent
\textbf{Test data.} The test data are collected in a similar fashion to the training/validation data. Experiments for 5 different samples in the same material group (C) are considered. Experiments are performed for different temperatures, exposure times and acquisition rates. Concentric ROIs with increasing $q$ are used. 10 ROIs with the smallest $q$-s are considered. However, no visual inspection of the data is done prior to model application and the ROIs with slow dynamics are not rejected. ROIs, where the full-range 1TCF fit to Eq.~\ref{eq:(3)} does not converge, are not considered. Overall, 12060 inputs (679 distinct targets) are considered in the test set. The distribution of the parameters from Eq.~\ref{eq:(3)} in shown in Fig.~\ref{fig:FigureS2}.
\noindent Unlike training/validation data, the contrast for normalization of test inputs is estimated from the 1TCF derived from the 2TCF) at lag=1 frame instead of the speckle visibility. This is done to reduce the computation time and to test the model performance for the cases when only the 50$\times$50 2TCFs, and not the scattered images, are available. No adjustment is done to the baseline as the input data does not provide a good estimate for it. Thus, for each of the noisy 2TCF$_{noisy}$, the model \emph{input} is calculated as:
\begin{equation}\label{eq:(5)}
input = (2TCF_{noisy} - 1)/1TCF_{noisy}(lag=1) + 1
\end{equation}
The denoised 2TCF$_{denoised}$ is then obtained from the output as: 
\begin{equation}\label{eq:(6)}
2TCF_{denoised} = (output - 1)*1TCF_{noisy}(lag=1) + 1
\end{equation}

\noindent 
\noindent\textbf{Non-equilibrium test.}
\noindent For the example of ageing dynamics considered in this work, the model is applied to each $50\times50$ piece of the raw 2TCF along its [1,1] diagonal with the step size 5 frames, starting at the first frame.
Prior the application of the model, each input is obtained from a raw 2TCF$_{noisy}$ as:
\begin{equation}\label{eq:(7)}
input = (2TCF_{noisy} - C_{\infty}^{*})/\beta^{*}(lag = 1) +1 
\end{equation}
where $\beta^{*}(lag = 1)$ is the estimation of contrast at lag=1 as the mean of 2TCF$_{noisy}(lag =1)$ for frames 250-300 and $C_{\infty}^{*}$ is the estimation of the baseline as the mean of 2TCF$_{noisy}$ at lags 270-300. The reverse transformation is applied to the model output.
The overlapping model outputs between the current and the previous steps are averaged. The values outside of the $50\times50$ diagonal sliding window are remained unchanged. The same procedure is repeated with the model sliding window moving from the last frame towards the first frame. The two results are averaged to reduce the dominating influence of the later dynamics over the earlier dynamics and vice versa. The loss of the temporal resolution due to convolution between the model and the raw signal is not significant for the considered case of slowly-evolving sample dynamics.

\noindent\textbf{Model Training Details.} The cost function used for training the models is the sum of the Mean Squared Error (MSE) between the target 2TCF and the models' output and the MSE between the respective 1TCFs (without lag=0):  
\begin{equation}\label{eq:(8)}
cost = \frac{1}{2500\cdot m}\sum_{k =1}^{m} ||x^{out}_{k} -x^{target}_{k}||_{2} + \frac{1}{49 \cdot m}\sum_{k =1}^{m} ||1TCF(x^{out}_{k}) -1TCF(x^{target}_{k})||_{2}
\end{equation}
where \(x^{out}_{k}\) is the model output for the $k$–th training example and \(x^{target}_{k}\) is the corresponding target’s pixel, \emph{m} is the number of examples, \(||\cdot||_{2}\) stands for \emph{2-norm}. 

\noindent At every training epoch, batches of size 8 are processed. Adam \cite{Kingma_2014} optimizer with initial learning rate from 2.5e-6 to 4e-5 is used. Learning rate is reduced by a factor of 0.9995 at every epoch. Initial weights in the convolutional and linear layers are assigned according to Xavier uniform initialization \cite{Glorot_2010}. The models are trained with Nvidia GPU accelerator GeForce RTX 2070 Super. For the selected CNN-ED configuration, the average training time is 27 seconds per epoch with 9-82 epochs necessary to train a model. Each input or target takes 30 kB of computer memory.

\noindent A model application does not require a GPU and, in fact, can be preformed faster without transferring the 2TCF data to a GPU. When using a CUDA accelerator, loading the model from a file, converting the 2TCF from numpy arrays to a CUDA Pytorch tensor, application of the model and converting the result back to a numpy array takes 2.3 ms on average with pure model computation taking 0.48 ms. Without using a CUDA accelerator, the corresponding times are 1.4 ms and 0.57 ms, respectively.  

\section*{Acknowledgements}

The authors thank A. Fluerasu and M. Fukuto for fruitful discussions. This research used CHX and CSX beamlines and resources of the National Synchrotron Light Source II, a U.S. Department of Energy (DOE) Office of Science User Facility operated for the DOE Office of Science by Brookhaven National Laboratory(BNL) under Contract No. DE-SC0012704 and under a BNL Laboratory Directed Research and Development (LDRD) project 20-038 "Machine Learning for Real-Time Data Fidelity, Healing, and Analysis for Coherent X-ray Synchrotron Data" 

\section*{Author contributions statement}

A.M.B, A.M.D, L.W and T.K. conceived the idea, L.W. performed the beamline experiments, generated the XPCS results, and identified individual scans for model development, T.K. processed the data for the model, A.M.D and T.K. wrote the code, M.R. provided technical consultation, T.K., L.W., A.M.D, A.M.B and M.R. analyzed the model performance, T.K. wrote the manuscript with contribution from all authors.

\section*{Additional information}

\noindent\textbf{Accession codes and Data availability} The code and the data used for the model training can be found at GitHub repository, \emph{https://github.com/bnl/CNN-Encoder-Decoder} at the time of publication. 

\noindent\textbf{Competing interests} The authors declare no competing interests.

\newpage

\noindent{\LARGE\bfseries Supplemental Materials for: Noise Reduction in X-ray Photon Correlation Spectroscopy with Convolutional Neural Networks Encoder-Decoder Models}

\section*{I.~Distribution of dynamics parameters for the training, validation  and test set}

\begin{figure}[ht]
\centering
\includegraphics[]{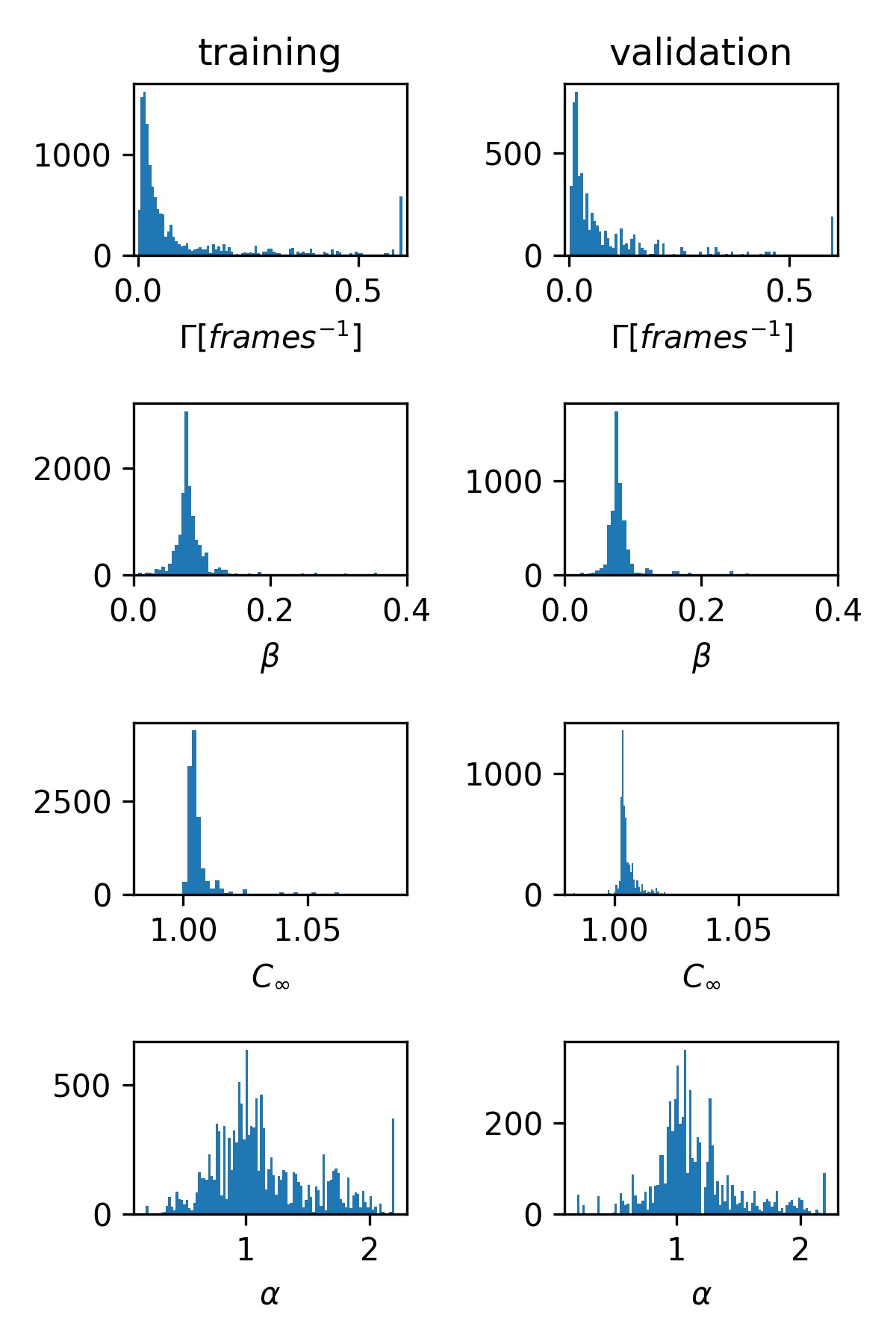}
\caption{Dynamics' parameters distribution for the training (left column) and validation (right column) sets.}
\label{fig:FigureS1}
\end{figure}

\textbf{Training and validation sets.} The distributions of the dynamics parameters for the train and validation sets are obtained by fitting the full-range 1TCF (calculated from all available frames) for each ROI of each experiment to the Eq.~3 and by sampling according to the number of inputs obtained from the corresponding 2TCF. For example, for an experimental dataset with 5 ROIs and 200 frames, each ROI gives seven \(50\times50\) inputs. We obtain 5 sets of parameters ($\beta$, $\Gamma$, $\alpha$ and $C_{\infty}$), one for each ROI. We then copy each parameter set to the training parameters pool 7 times. The distributions of the parameters in the training and the validation pools are shown in Fig.~\ref{fig:FigureS1}.

\begin{figure}[ht]
\centering
\includegraphics[]{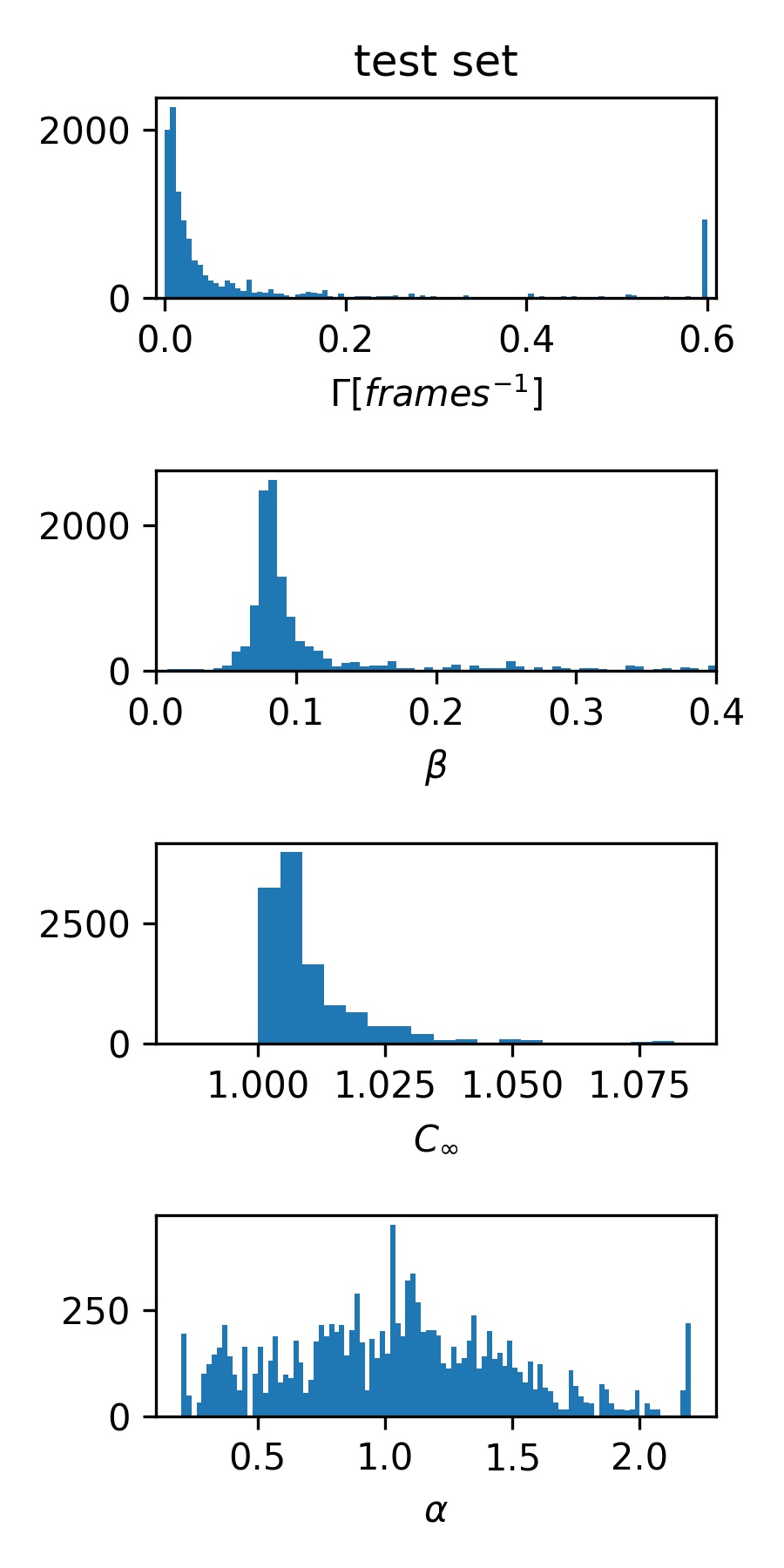}
\caption{Dynamics' parameters distribution for the test set.}
\label{fig:FigureS2}
\end{figure}

\noindent\textbf{Test set.} The test set contains data from similar experiments as used for training/validation. Each experiment has 600-1000 frames. This set is used for establishing applicability limits of the model. While the 1TCFs for these data are not perfectly described by Eq.~3, they can be approximated with it. We obtain the distributions for the dynamics parameters in a similar way as for the training/validation data. The distributions are shown in Fig.~\ref{fig:FigureS2}. 

\section*{II.~Model performance on the validation set}

\begin{figure}[pt]
\centering
\includegraphics[]{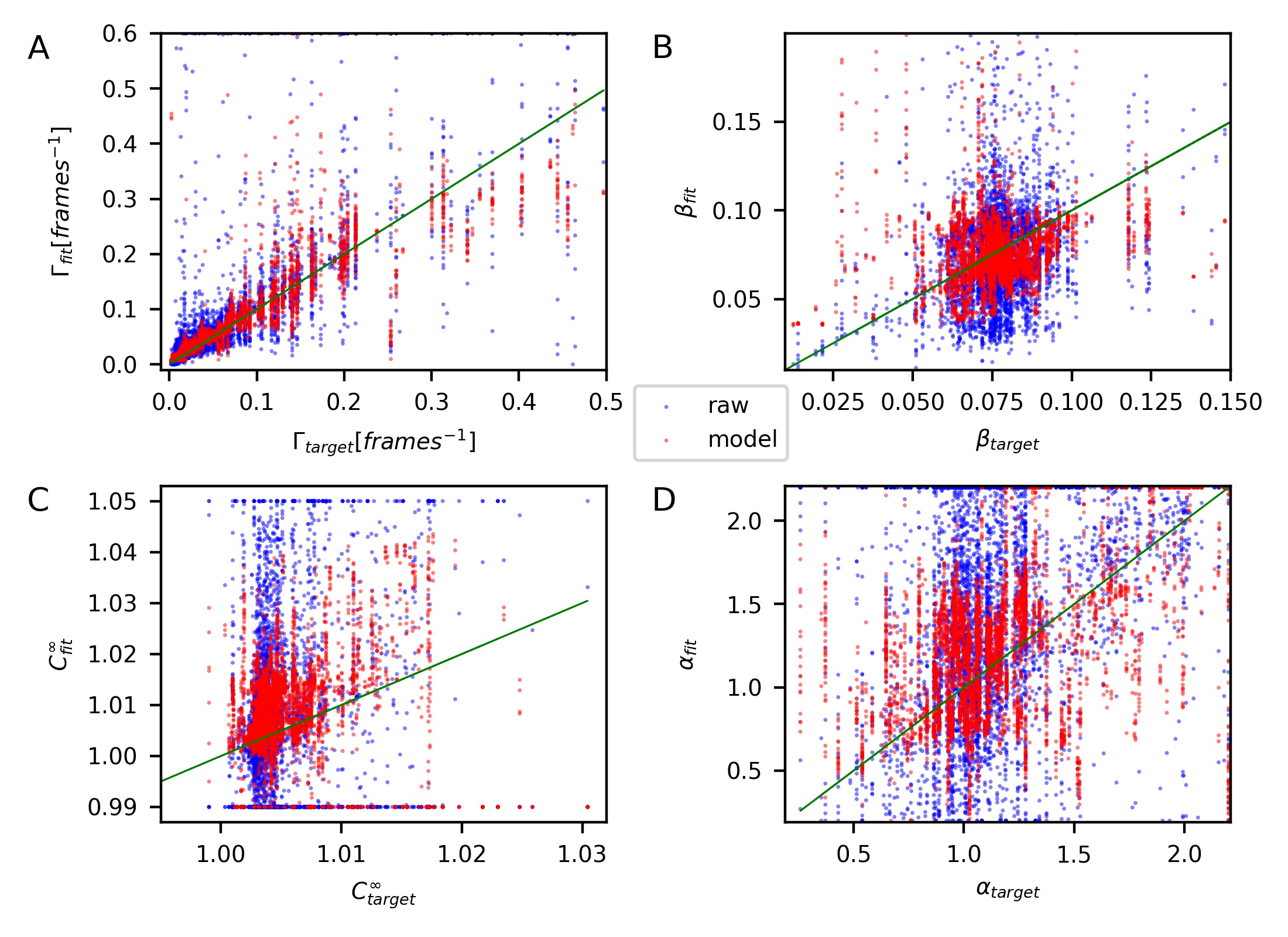}
\caption{Comparison of dynamics parameters extracted for raw (blue) and denoised (red) 50$\times$50 2TCF from the validation set. Horizontal axes reflects the values extracted from the corresponding target 2TCF. The green lines with slope 1 are included for convenience to represent a full correlation between the target and the fit.}
\label{fig:FigureS3}
\end{figure}

\noindent We check the performance of the best ensemble (10 models) on the validation set by comparing the errors of dynamics parameters (Eq.~~\ref{eq:(3)}) extracted from the raw input 2TCFs and the denoised outputs. The error are calculated with respect to the corresponding parameters, extracted from the raw full-range (target) 2TCF. The comparison is shown in Fig.~\ref{fig:FigureS3}. One can see that the rate $\Gamma$ is extracted from the ensemble's output with a good precision for $\Gamma < 0.2 frames^{-1}$ (the contrast drops by half in 2 or more frames). Above $\Gamma=0.2 frames^{-1}$, the variance for the $\Gamma$ extracted from the denoised data is similar to the one of the $\Gamma$ extracted from the raw data. Other dynamics parameters are generally extracted with better precision from the denoised data than from the raw data. Moreover, for some of the raw inputs, the fit to the Eq.~~\ref{eq:(3)} does not converge within the reasonable parameters bounds, but the corresponding model outputs can be fit within the same bounds. Note, the precision  of the $\beta$ is largely dependent on the accuracy of extracting the speckle visibility from the photon distribution in a single frame, which serves as the normalization parameter. 

\noindent The comparison is summarized in Table~~\ref{tab:TableS1}. It includes the mean square error (MSE) of 1TCF, the mean absolute relative error (MRE) for the $\Gamma$ and the MSE for all other parameters. It is important to consider the relative errors for the $\Gamma$ instead of absolute errors since its values can be very close to zero and a small absolute error may still be significant.

\noindent Note, that performance of the model is optimized to the validation set via the early stopping criteria and that the good accuracy is expected for the set. The model's generalizability and application limits are determined below based on the test examples, which are not accessed during the training.

\begin{table}[ht]
\centering
\begin{tabular}{|l|l|l|l|}
\hline
 & raw 50$\times$50 & model output  \\
\hline
do not converge & 379 & 0   \\
\hline
MSE(1TCF) & 2.8e-5 & 1.05e-5   \\
\hline
MRE $\Gamma$ & 0.72 & 0.46   \\
\hline
MSE $\beta$ & 0.0045 & 0.0016   \\
\hline
MSE $C_{\infty}$ & 0.00027 & 0.00009   \\
\hline
MSE $\alpha$ & 0.45 & 0.16\\
\hline
\end{tabular}
\caption{\label{tab:TableS1}Comparison of parameters' errors extracted from the raw input data and from the corresponding model's output for the validation set (5449 examples).}
\end{table}

\section*{III.~Model performance on the test set}
We check the performance of the best ensemble (10 models) on the test set by comparing the errors of dynamics parameters (Eq.~~\ref{eq:(3)}) extracted from the raw input 2TCFs and the denoised outputs with respect to the corresponding parameters extracted from the raw full-range (target) 2TCF. Unlike the training/validation sets that only include the 2TCFs with noticeable de-correlation within the first 50 frames, the test set includes all available 2TCFs. There are several difficulties for comparison when all four parameters in the Eq.~3 are relaxed: 
\begin{itemize}
  \item Sample dynamics are more complicated than the Eq.~~\ref{eq:(3)} and may contain several exponential terms and/or terms with variable parameters.
  \item The considered 49 time points might not be enough to unequivocally distinguish between two or more possible dynamics with different parameters, especially in cases of very slow dynamics.
\end{itemize}

\begin{figure}[ht]
\centering
\includegraphics[width=\linewidth]{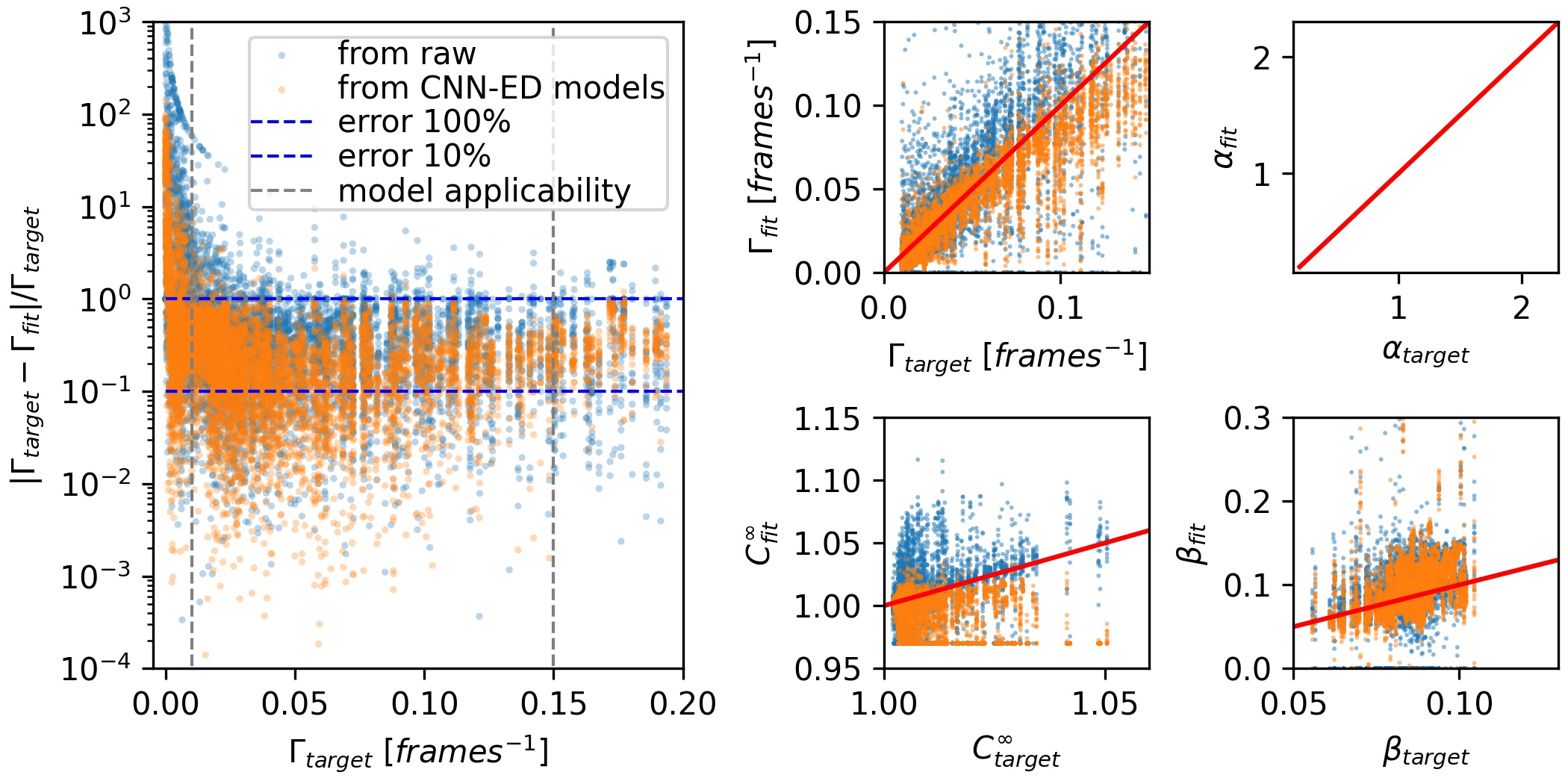}
\caption{Comparison of accuracy of extracting the dynamics' parameters with $\alpha =1$ for raw (blue) and denoised (orange) 50$\times$50 2TCFs from the test set. The left panel shows the relative error of $\Gamma$ versus underlying values of $\Gamma_{target}$ (extracted from full-sized 2TCF). The vertical dashed lines show the lower and the upper boundaries of the model applicability. The right panel show the values for each of the dynamics parameters for the inputs within the model applicability limits versus the values extracted from the corresponding target 2TCF.}
\label{fig:FigureS4}
\end{figure}

\noindent The goal of this work is not to correctly characterize dynamics of the materials used for experiments in the test set, but to identify how close the outputs of the model are to the respective correlation functions measured with good statistics (the target). Thus, for a more deterministic comparison, we reduce the number of parameters and fix \(\alpha\) to 1. Fixing or highly restricting the fit parameters is a common practice in XPCS analysis. The results are shown in Fig.~\ref{fig:FigureS4}. 

\noindent Firstly, we set the applicability limits of the model by identifying the regions where the relative error for $\Gamma$ is too large. The distribution of relative errors of $\Gamma$ for different target values of $\Gamma_{target}$ indicates that the ensemble does not perform well for small values of $\Gamma < 0.01 frames^{-1}$. At such rates, only a portion of the dynamics is complete within the first 50 frames: the contrast drops by 60 or less percent at furthest available (50$^{th}$) frame. Above $\Gamma = 0.01 frames^{-1}$ the relative error of $\Gamma$ is generally below 100\%, which can be an acceptable level depending on the experiment. We thus select \(\Gamma = 0.01 frames^{-1}\) as the lower bound for the model applicability range. We select \(\Gamma = 0.15 frames^{-1}\) as the upper bound since above this value a better precision can be achieved from the raw outputs than from the model. The upper bound is smaller than for the validation set, which partially can be explained by errors in estimation of contrast used for the input normalization. For the test set, the normalization parameter is estimated from the first off-diagonal elements of the input $50\times50$ 2TCF, while for the training and validation sets it is estimated from all available single frames using speckle visibility spectroscopy. Errors in contrast estimation can be especially pronounced for the cases of fast dynamics, where the contrast drops significantly within a a single frame.

\begin{figure}[ht]
\centering
\includegraphics[width=\linewidth]{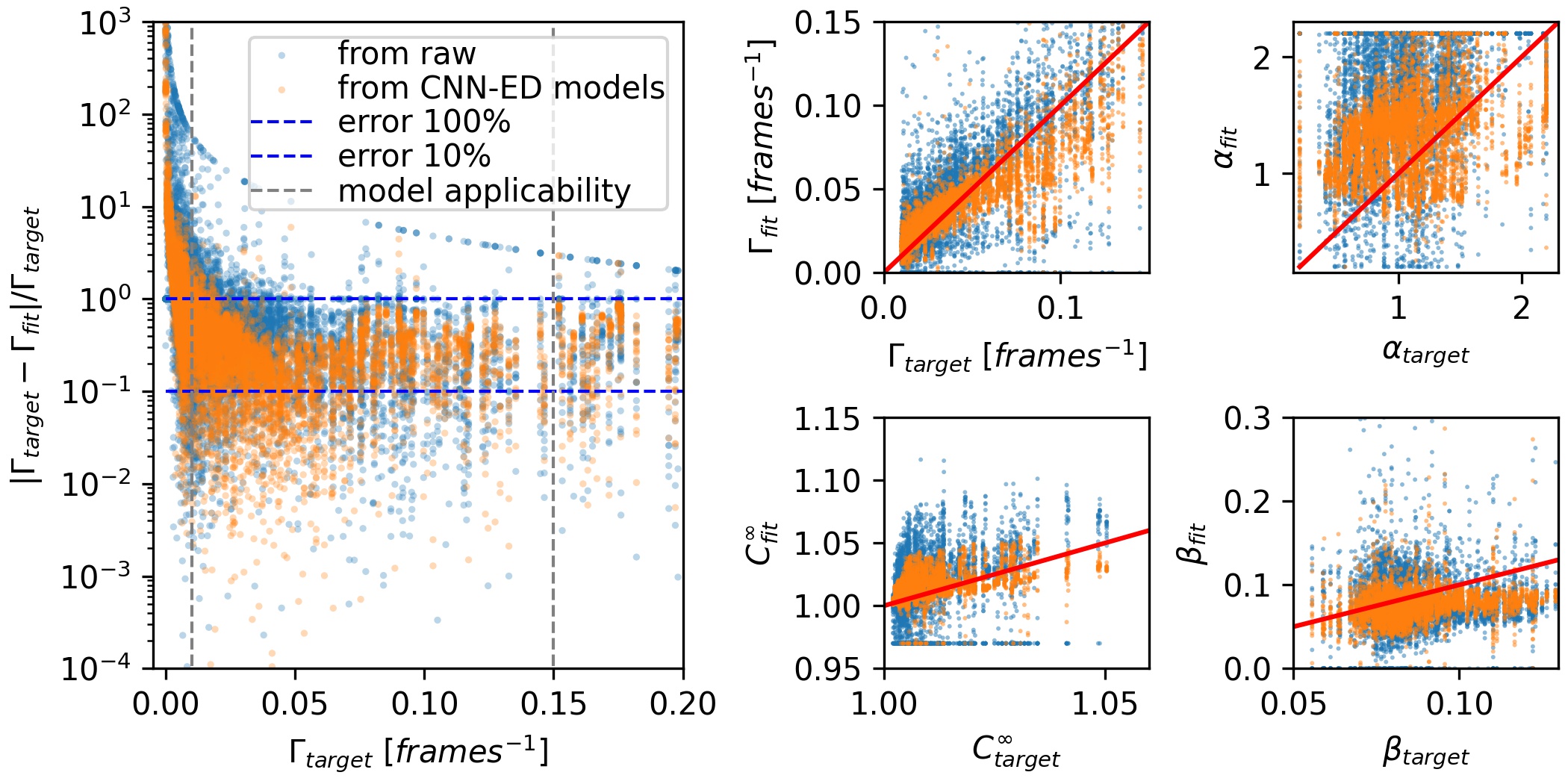}
\caption{Comparison of accuracy of extracting the dynamics' parameters with the relaxed parameter $\alpha$ for raw (blue) and denoised (orange) 50$\times$50 2TCFs from the test set. The left panel shows the relative error of $\Gamma$ versus underlying values of $\Gamma_{target}$ (extracted from full-sized 2TCF). The vertical dashed lines show the lower and the upper boundaries of the model applicability. The right panel show the values for each of the dynamics parameters for the inputs within the model applicability limits versus the values extracted from the corresponding target 2TCF.}
\label{fig:FigureS5}
\end{figure}

\noindent The comparative accuracy for the other dynamics parameters are also shown in Fig.~\ref{fig:FigureS4}. The accuracy for measuring amplitude \(\beta\) is similar for the raw data and for the model output, which is expected since the normalization for the model inputs is  based on the contrast measured for the raw data. However, the baseline is under-fit for the large values, that are not present enough in the training set. 
For comparison, Figure~\ref{fig:FigureS5} shows the same dependencies for the case when the \(\alpha\) is not fixed. 

\noindent The summary statistics for the Fig.~\ref{fig:FigureS4}and Fig.~\ref{fig:FigureS5} are in the Table~\ref{tab:TableS2}. Only the cases with $0.01 frames^{-1}< \Gamma_{target} < 0.15 frames^{-1}$ are included.

\begin{table}[ht]
\centering
\begin{tabular}{|l|l|l|l|}
\hline
 & raw 50$\times$50 & model output  \\
\hline

MSE(1TCF) & 0.012 & 6.6e-5   \\
\hline
\hline
MRE $\Gamma$, $\alpha=1$ & 0.56 & 0.28   \\
\hline
MSE $\beta$, $\alpha=1$ & 0.001 & 0.0007   \\
\hline
MSE $C_{\infty}$, $\alpha=1$ & 0.0007 & 0.00065   \\
\hline
do not converge, $\alpha=1$ & 415 & 18   \\
\hline
\hline
MRE $\Gamma$ & 0.76 & 0.33   \\
\hline
MSE $\beta$ & 0.0042 & 0.0011   \\
\hline
MSE $C_{\infty}$ & 0.0006 & 0.0002   \\
\hline
MSE $\alpha$ & 0.69 & 0.24\\
\hline
do not converge & 343 & 6   \\
\hline
\end{tabular}
\caption{\label{tab:TableS2}Comparison of parameters errors extracted from the raw test data (7228 examples) and from the corresponding model outputs.}
\end{table}

\section*{IV.~Model performance at different noise levels}
\begin{figure}[ht]
\centering
\includegraphics[]{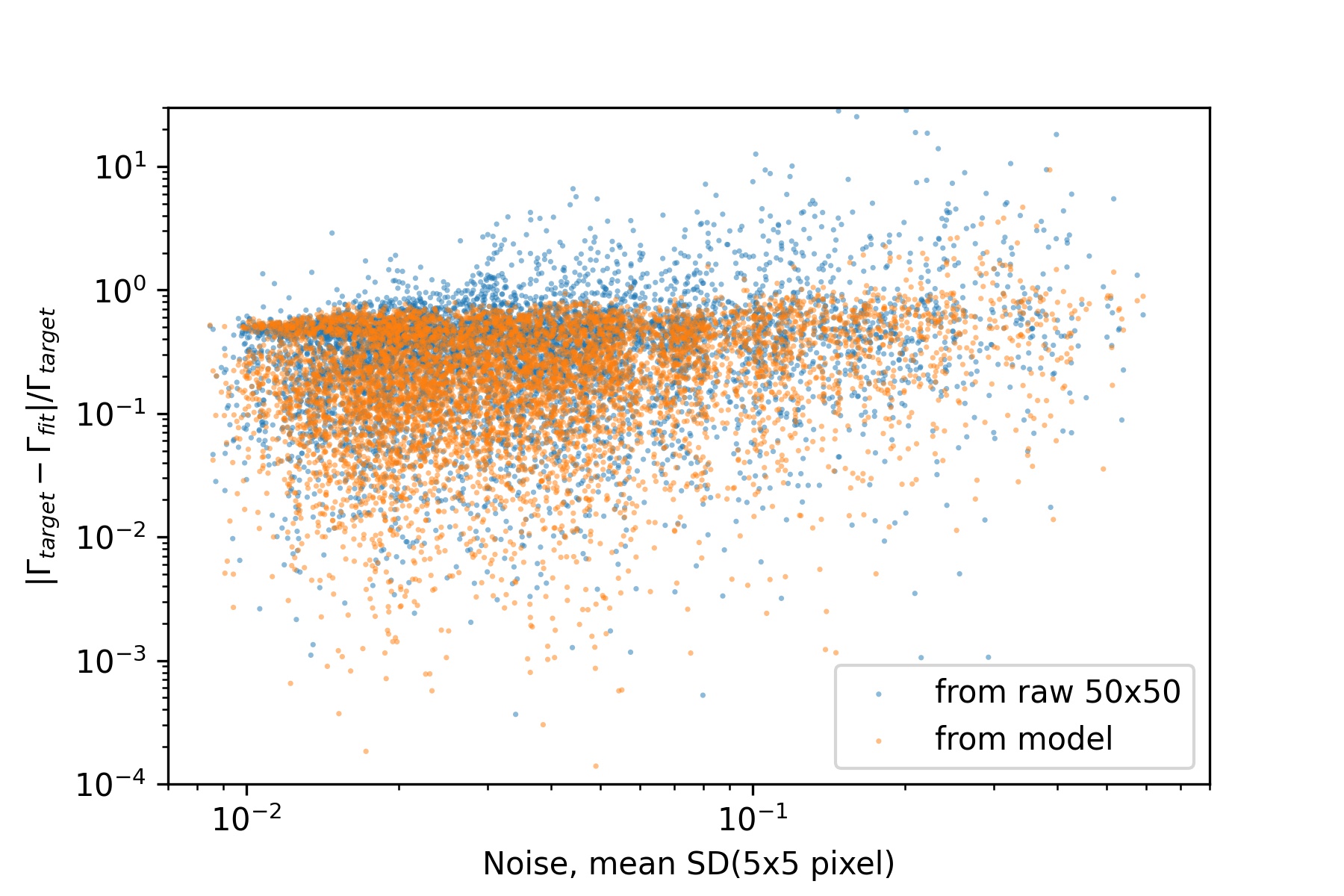}
\caption{Relative error of $\Gamma$ as a function of 2TCF noise. }
\label{fig:FigureS6}
\end{figure}

Above, we have identified the limiting cases of rates of sample dynamics, beyond which the model does not perform well. It is expected that the noise level of input 2TCFs can limit the model accuracy as well. As the measure of noise we use the average standard deviation of pixel values within $5\times5$ fields of an input 2TCF. To calculate the noise, a 2TCF image is split into $5\times5$ pixels regions, and a standard deviation of intensity is calculated for each region. Then, the values are averaged among all the regions. The values are calculated prior to normalization of contrast to 1 (typical contrast values are 0.06-0.1). We compare $MRE(\Gamma)$  for the raw input data and the model outputs from the Fig.~\ref{fig:FigureS5} as functions of the input noise. The comparison is shown in Fig.~\ref{fig:FigureS6}. Only the examples within the $0.01 frames^{-1} <\Gamma_{target}<0.15 frames^{-1}$ are included. The accuracy of parameters extracted from both the raw data and from the model output deteriorates at larger noise. However, in a wide range of input noise the accuracy of the model's output is better than the raw 2TCF values as it does not have any extreme errors. Above a certain noise level (around 0.2), the $MRE(\Gamma)$ is comparable for both model's output and the raw data.

\begin{figure}[ht]
\centering
\includegraphics[]{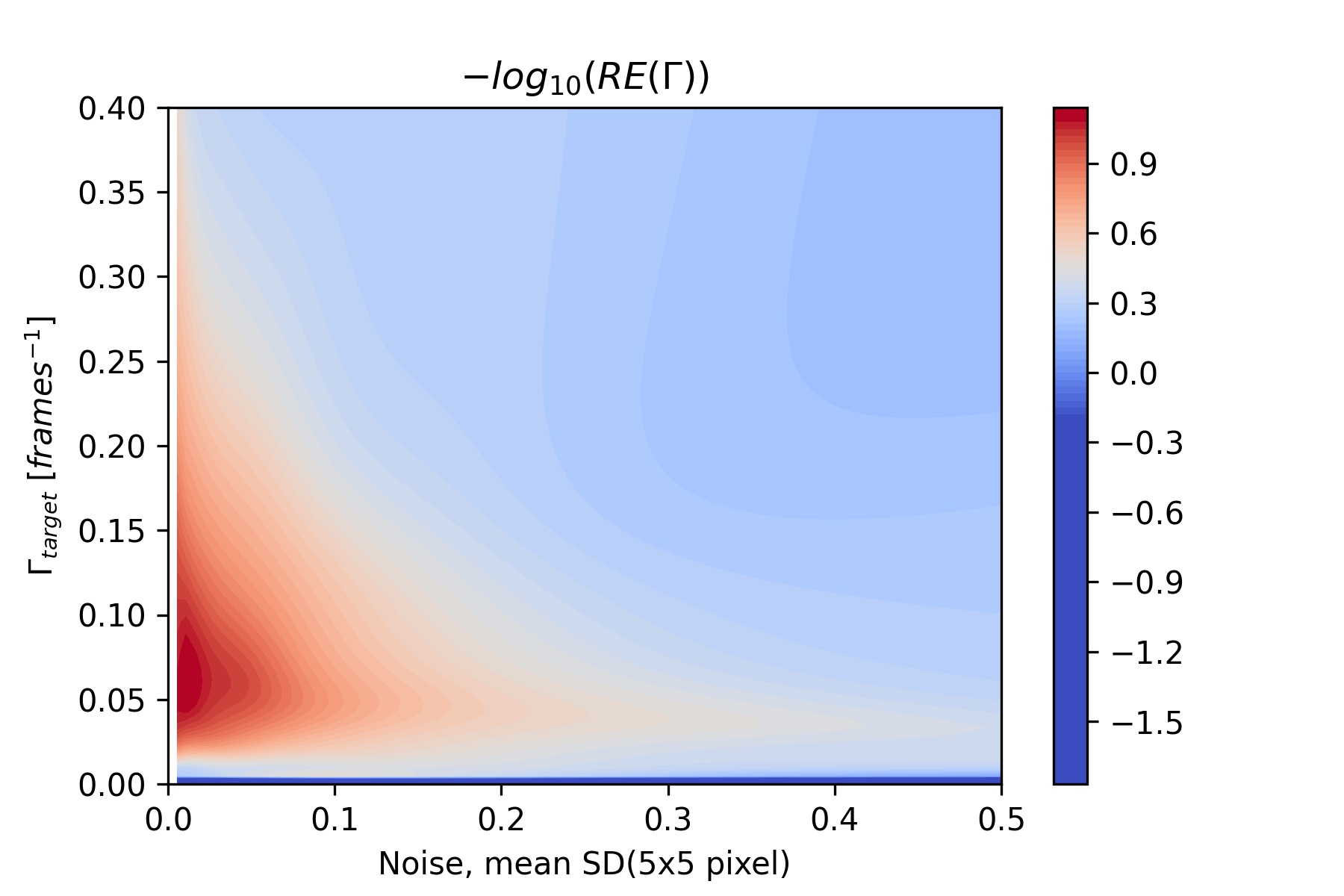}
\caption{The contour plot of the logarithm of relative error for $\Gamma$ as a function of underlying values of $\Gamma_{target}$ and the noise level in 2TCF (the mean of standard deviations withing 5$\times$5 pixel fields). High precision (low error) is observed for low-noise inputs for dynamics with $0.01 frames^{-1} <\Gamma<0.15 frames^{-1}$ and low precision is observed for inputs with large noise and/or extreme dynamics rates. A log-normal kernel smoothing is used for interpolation and reducing the scatter.}
\label{fig:FigureS7}
\end{figure}

\noindent The relative error of $\Gamma$ extracted from the model outputs is shown in Fig.~\ref{fig:FigureS7} as a function of the underlying rate of sample's dynamics $\Gamma_{target}$ (as in Fig.~\ref{fig:FigureS4}) and the noise level of the input 2TCF. All tests examples are included.

\section*{V.~Analysis of CNN-ED layers}
\begin{figure}[ht]
\centering
\includegraphics[width=\linewidth]{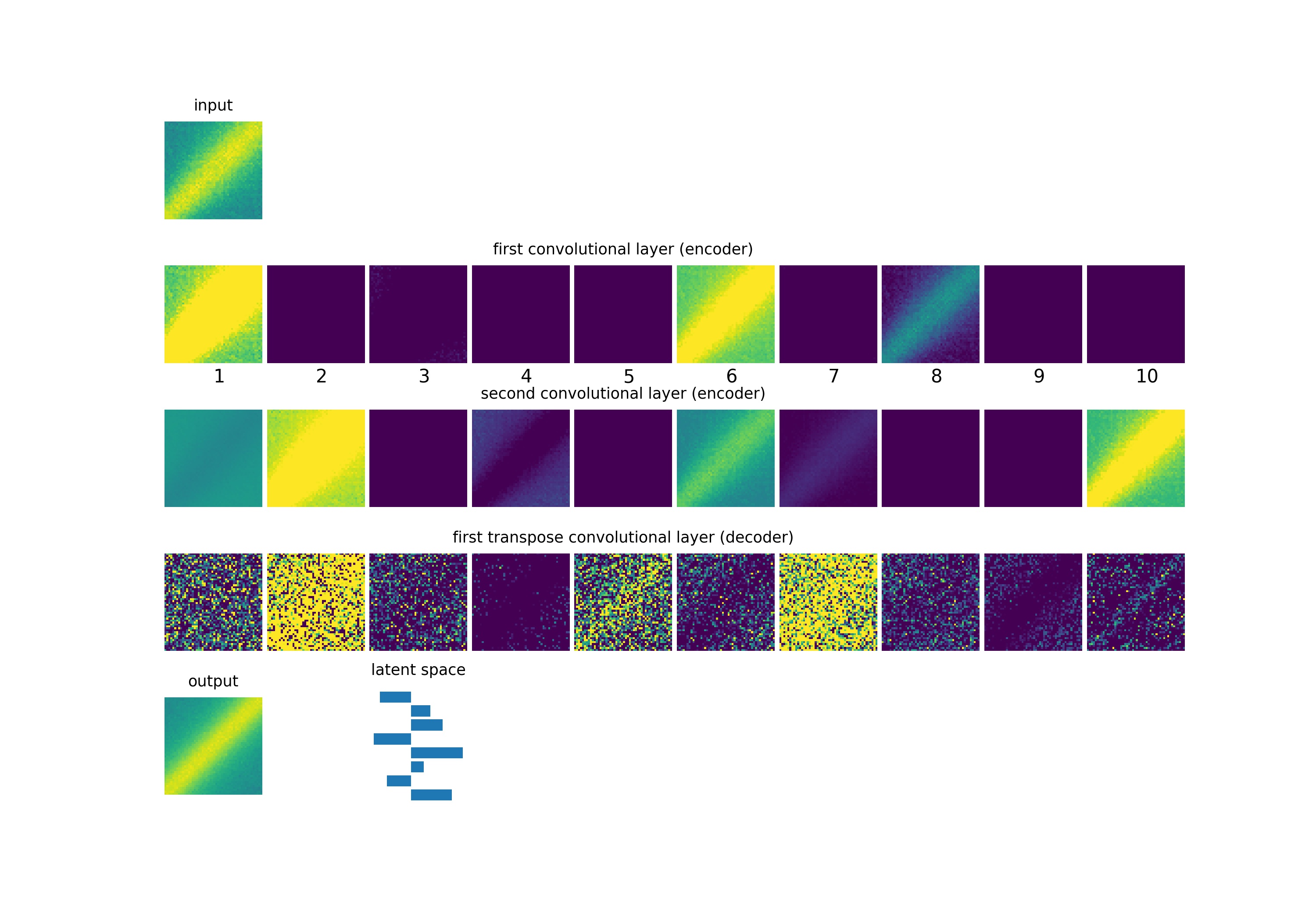}
\caption{Outputs of each layer of an CNN-ED model for the input with high signal-to-noise ratio. In the first encoder layer channels 2,3,4,5,7,9 and 10 are not activated for this input. In the second layer, channels 3, 5, 8 and 9 are not activated. All of these channels are activated for a noisy example, shown in Fig.S8.}
\label{fig:FigureS8}
\end{figure}

\begin{figure}[ht]
\centering
\includegraphics[width=\linewidth]{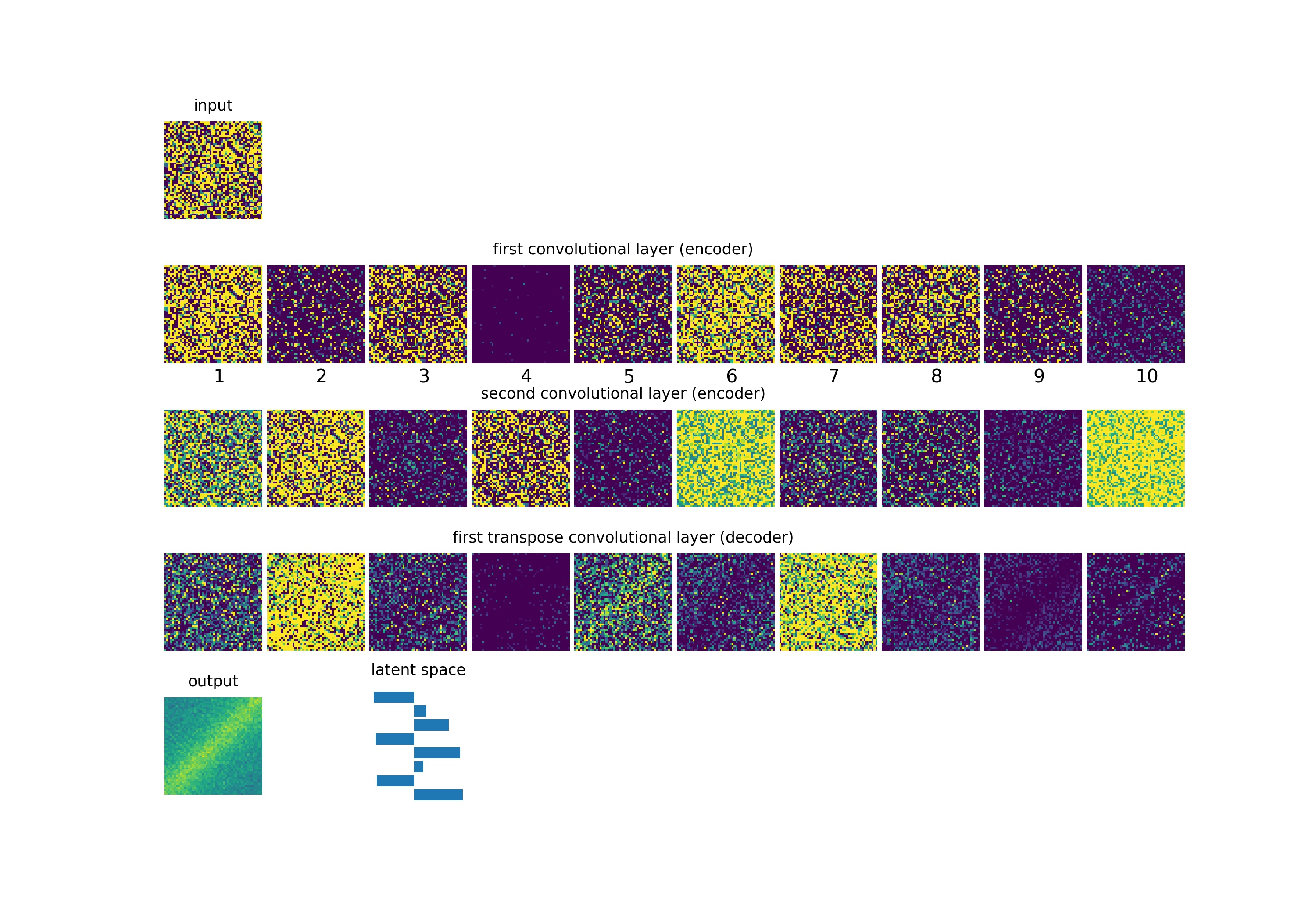}
\caption{Outputs of each layer of an CNN-ED model for the input with low signal-to-noise ratio.}
\label{fig:FigureS9}
\end{figure}

To better understand how a CNN-ED model handles different levels of noise, we look at the activation maps of its layers for inputs with low (Fig.~\ref{fig:FigureS8}) and high (Fig.~\ref{fig:FigureS9}) noise. For the case of the low noise input, not all channels in the encoder layers are activated. However, for the case of the high noise input, all channels are activated, meaning that some channels are meant for handling the extreme pixel values. The latent space variables are generally non-zero for all types of inputs. 

\begin{figure}[ht]
\centering
\includegraphics[width=\linewidth]{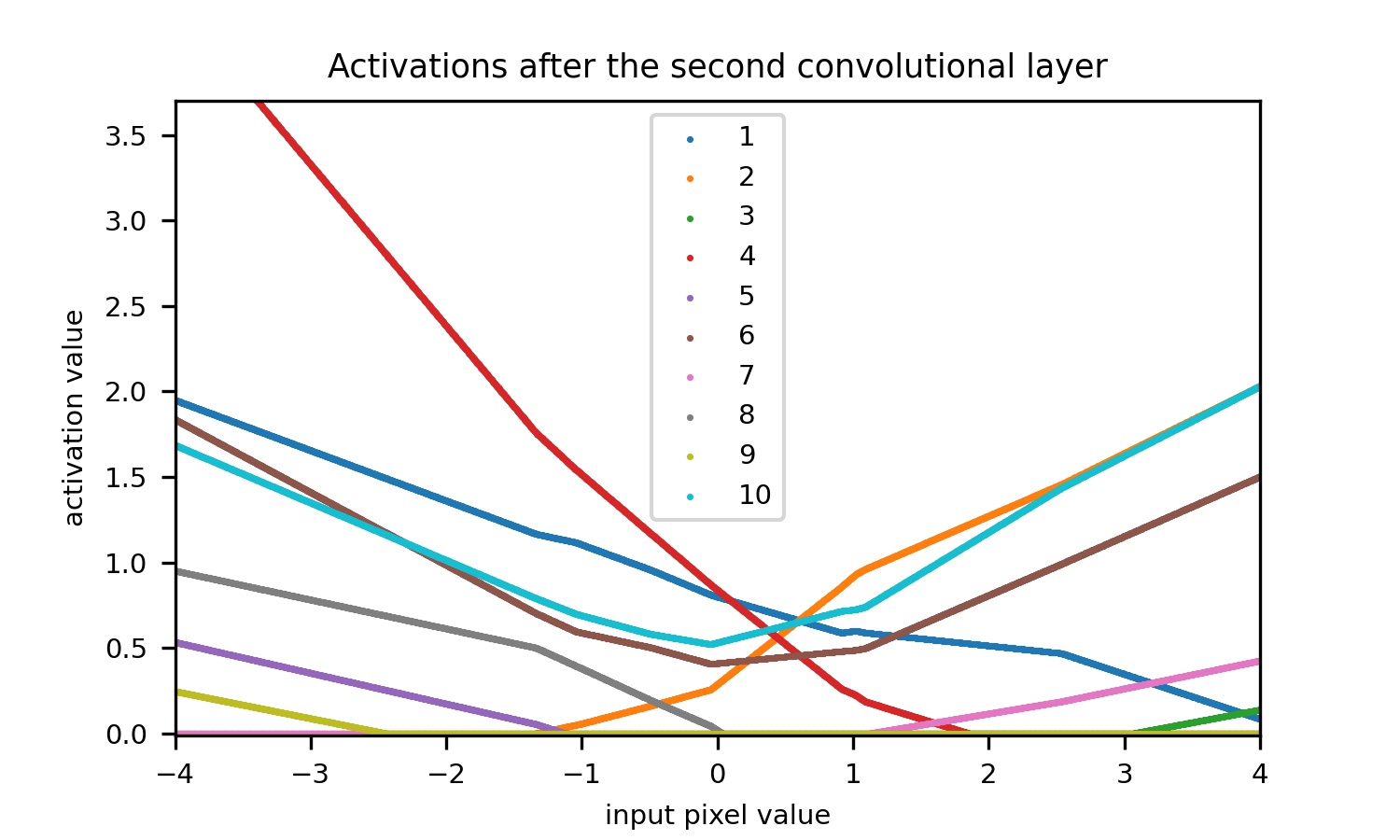}
\caption{10 activation functions of the encoder for one of the CNN-ED models.}
\label{fig:FigureS10}
\end{figure}

\noindent Due to the nature of the 2TCF and the scaling of the inputs, the “good” values of the inputs and outputs should be between 1 and 2. To see how different pixel values are transformed during the encoding stage, we look (Fig.~\ref{fig:FigureS10}) at outputs of the second encoding layer after application of the \emph{ReLU} function. Models trained with different random weight initialization have different nonlinear activation functions. The activation functions of the encoder for the model in Fig.~\ref{fig:FigureS8} and Fig.~\ref{fig:FigureS9} are shown in Fig.~\ref{fig:FigureS10}. One can see that some layers (e.g. 5, 8, 9) are only activated at small (negative) correlation values, which are the noise since they are unnatural for the correlation function according to the Eq.~~\ref{eq:(1)}.

\section*{VI.~Model regularization}
To avoid over-fitting, several model regularization measures are implemented:
\begin{itemize}
  \item selecting the simplest model configuration from those that give similar performance on the validation set
  \item restricting the latent space dimension
  \item having noise in both inputs and targets
  \item augmenting the data
  \item early stopping while training
  \item combining several models trained with different random initialization into an ensemble
\end{itemize}

\begin{figure}[ht]
\centering
\includegraphics[]{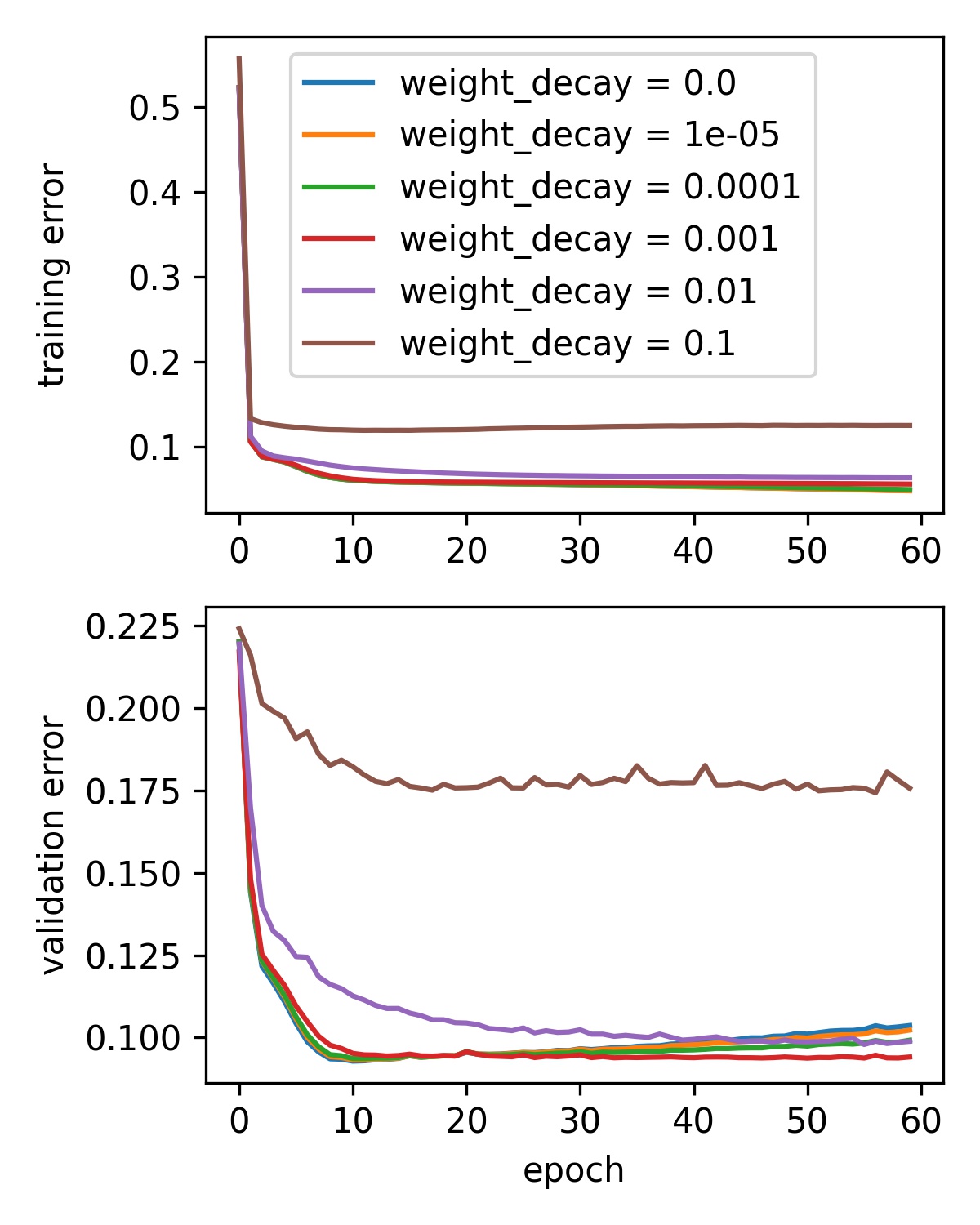}
\caption{Model training with different \emph{'weight\_decay'} parameters.}
\label{fig:FigureS11}
\end{figure}

\noindent We have also attempted introducing additional weight regularization via increasing the \emph{'weight\_decay'} parameter of the Adam optimization algorithm and using dropout channels in the convolutional layers. Both approaches did not improve the model performance on the validation set. Evolution of the training and validation cost functions during a model (10 channels in both convolutional layers, latent space dimension is 20) training with different \emph{'weight\_decay'} is shown in Fig.~\ref{fig:FigureS11}. While the increased values of the \emph{'weight\_decay'} lead to reduced variance of the model by flattening the tail of the validation cost, they do not reduce the minimum value of it. Similar situation is observed for increasing the dropout probability for the convolutional layers. Thus, we decided not to employ the additional regularization approaches.

\end{document}